\documentclass{aastex63}
\usepackage[utf8]{inputenc} 
\received{June 1, 2019}
\revised{January 10, 2019}
\accepted{\today}
\submitjournal{ApJ}

\shorttitle{Delta Gravity: CMB and supernovas}
\shortauthors{San Martín et al.}
\graphicspath{{./}{figures/}}

\usepackage{amsmath}

\begin{document}

\title{Observational constraints in Delta Gravity: CMB and supernovas}

\correspondingauthor{Marco San Martín}
\email{mlsanmartin@uc.cl}

\author{Jorge Alfaro}
\affiliation{Instituto de Física, Pontificia Universidad Católica de Chile, Avda. Vicuña Mackenna 4860, Santiago 7820436, Chile}

\author[0000-0003-2565-3174]{Marco San Martín H.}
\affiliation{Instituto de Astrofísica, Pontificia Universidad Católica de Chile, Avda. Vicuña Mackenna 4860, Santiago 7820436, Chile}

\author[0000-0001-6739-5746]{Carlos Rubio}
\affiliation{Instituto de Física, Pontificia Universidad Católica de Chile, Avda. Vicuña Mackenna 4860, Santiago 7820436, Chile}

\nocollaboration{3}





\begin{abstract}

Delta Gravity is a gravitational model based on an extension of General Relativity given by a new symmetry called $\tilde{\delta}$. In this model, new matter fields are added to the original matter fields, motivated by the additional symmetry. We call them $\tilde{\delta}$ matter fields. This model predicts an accelerating Universe without the need to introduce a cosmological constant. In this work, we study the Delta Gravity prediction about the scalar CMB TT power spectrum using an analytical hydrodynamical approach. To fit the Planck satellite's data with the DG model, we used a Markov Chain Monte Carlo analysis. We also include a study about the compatibility between SNe-Ia and CMB observations in the Delta Gravity Context. Finally, we obtain the scalar CMB TT power spectrum and the fitted parameters needed to explain both SNe-Ia Data and CMB measurements. The results are in a reasonable agreement with both observations considering the analytical approximation. We also discuss if the Hubble Constant and the Accelerating Universe are in concordance with the observational evidence in the Delta Gravity context.

\end{abstract}



\section{Introduction} \label{sec:intro}

Cosmology is a very active area of study, where many observational data allow a better understanding of theoretical physics. The scientific community has evidence that most of the composition of the Universe is unknown. This sector comprises two kinds of components called dark matter (DM) and dark energy (DE) (\cite{weinberg2008cosmology, Riess1998, Perlmutter_1999, DM_DE4, Planck2018}).
The DM was initially detected by \cite{Zwicky_1937}, when he observed that some clusters were not principally made of stars or clusters of stars, but consists predominantly of matter which does not emit light. Then \cite{Rubin_1970,Rubin_1980}  found that the DM is the principal component of galaxies in terms of mass. Today we know that DM dominates the galaxies (\cite{Beasley_2016ApJ}) and the cosmological simulations such as \cite{Millenium_project,Vogelsberger_illustris,Vogelsberger2020,Wang_2020} show that DM plays an essential role as a source of the gravitational potential.\medskip

Regarding the DE, this is the main component of the Universe, and it is strictly necessary to reproduce the Universe's acceleration in the standard cosmological model called $\Lambda$CDM (\cite{Riess_1998,Perlmutter_1999,DM_DE4,Planck2018}). Despite the observational evidence, the origin of the DE in the Einstein field equations or in the Einstein-Hilbert action is no clear (\cite{DarkEnergyPaper}). In early times after the Big Bang, this constant is irrelevant, but at the later stages of the evolution of the Universe, $\Lambda$ will dominate the expansion, explaining the acceleration. Such small $\Lambda$ ,which is commonly associated to the vacuum energy, is very difficult to generate in quantum field theory models because the predictions reach up to 120 orders of magnitude far from the observed $\Lambda$ in cosmology (\cite{DM_DE5},\cite{DarkEnergyPaper}). Moreover, in other attempts to obtain a better value, the result is about 54 orders of magnitude far from the $\Lambda$ observed value (\cite{DarkEnergyPaper}). This explanation is not satisfactory.\medskip

In the last decades, there have been various proposals to explain the observed acceleration of the Universe. They involve the inclusion of some additional fields in approaches like Quintessence, Chameleon, Vector Dark Energy or Massive Gravity; The addition of higher-order terms in the Einstein-Hilbert action, like $f(R)$ theories and Gauss-Bonnet terms, and the introduction of extra dimensions for a modification of gravity on large scales (\cite{DE3}). Other interesting possibilities, are the search for non-trivial ultraviolet fixed points in gravity (asymptotic safety, \cite{GR_Weinberg}) and the notion of induced gravity (\cite{induced_gravity1,induced_gravity2,induced_gravity3,induced_gravity4}). The first possibility uses exact renormalization-group techniques (\cite{Ren_group_1}-\cite{Ren_group_4}) together with lattice and numerical techniques such as Lorentzian triangulation analysis (\cite{Lorentz_triang}). Induced gravity proposes that gravitation is a residual force produced by other interactions.\medskip

The CMB Planck's data and its power spectrum provide important information to fit many cosmological parameters (\cite{Planck2018}). These cosmological fluctuations have been deeply studied and numerically solved in programs such as CMBFast (\cite{cmbfast1, cmbfast2}) or CAMB  (\cite{camb2}). From the CMB observations and the SNe-Ia data, the $\Lambda$CDM model indicates that the Universe is composed by about 68\% of DE (\cite{ Planck2018}). \medskip

The State-of-the-art of cosmology is controversial. A measurement about the $H_0$ by \cite{Sorce_2012} found a value of $75.2 \pm 3.0$ km/(Mpc s). A few years later, \cite{Riess2016} found an observed value $H_0 = 73.24\pm 1.74 $ km Mpc$^{-1}$ s$^{-1}$ using new parallaxes from Cepheids. This measurement is important because it is independent from cosmological models. This value is 3.4 $\sigma$ higher than $66.93\pm 0.62$ km Mpc$^{-1}$ s$^{-1}$ predicted by $\Lambda$CDM with Planck. But the discrepancy reduces to 2.1 $\sigma$ relative to the prediction of $69.3\pm 0.7$ km Mpc$^{-1}$ s$^{-1}$ based on the comparably precise combination of
WMAP+ACT+SPT+BAO observations. This value was updated in \cite{Riess2018} using more precise parallaxes for Cepheids. The $H_0$ updated value at 2018, is $73.52
\pm 1.62$ km Mpc$^{-1}$ s$^{-1}$. All the results from \cite{Riess2016, Riess2018, Riess2019}  are incompatible with \cite{Planck2018}. This tension between both observations has been widely discussed. For instance, other researchers used methods independent of distance ladders and the CMB, and they found that the Hubble constant exceeds the Planck's results \cite{Pesce_2020,Suyu_2013}. However, the errors calculated in the local measurements of the $H_0$ have been criticized \cite{Efstathiou2014,Zhang_2017}. Other measurements based on the tip of the red giant branch (TRGB) have found that $H_0$ is close to $69.6 $ km/(Mpc s) (\cite{Freedman_2019,Freedman_2020}). By the other hand, \cite{Cardona_2017,Follin_2018} confirmed a high $H_0$ value and recently \cite{Wong_2020} used lensed quasars and found $H_0=73.3$ Mpc/(km s), which agrees with local measurements but tension with Planck observations.\medskip 

Many solutions have been proposed to explain this tension, such as extended models based on $\Lambda$CDM (\cite{Guo_2019}), time-varying DE density models  (\cite{risaliti_lusso_2019}), or cosmography models (\cite{Benetti_2019}). Others attempt modifications in the early-time physics, including a component of dark radiation (\cite{Bernal_2016}) or  analyzing early physics related to the sound horizon (\cite{Aylor_2019}). Many efforts related to the recombination physics have been developed to solve the Hubble tension (\cite{agrawal_2019,Lin_2019,Knox_2020}).\medskip

This controversy opens a window for new alternative theories based on modifications or variations of $\Lambda$CDM such as \cite{Camarena2018,Huang2016,Li_2013,Cede_o_2019,Xu_2019,Deser_2019,Anagnostopoulos_2019,PhysRevD.97.123504}, other proposals introduce modifications in the physics of neutrinos, for example \cite{Battye2014,Zhang_2014,Bernal_2016,Valentino_2016,Guo_2017_0,Feng2017,Zhao_2017,Guo_2017_1,Benetti_2017,Feng2018,Zhao2018,Benetti_2018,RoyChoudhury2019,Carneiro2019,Nakamura_2019} and others consider that DE can couple with DM: \cite{PhysRevD.88.023531,PhysRevD.89.103531,PhysRevD.96.123508,PhysRevD.96.043503,FENG2019100261,Yang_2018}.

Some independent studies support the idea that the tension is due more
to the physics rather than observational errors \cite{Benetti_2019, Bonvin_2016, Abbott_2018, Lemos_2019}. Others have found tension in the CMB analysis (\cite{Addison_2016,curvature}) or suggest errors  in the values predicted by Planck CMB (\cite{Spergel_2015}). Also, it has been suggested as a solution to include modifications in the Planck analysis through more free parameters and varying the Equation of State of DE \cite{DIVALENTINO2016242,PhysRevD.96.023523}  \medskip

The Delta Gravity (DG)  model (\cite{DeltaGravityL}) emerges as a model of gravitation that is very similar to classical GR but could make sense at the quantum level.  DG could give clues about some incompatibilities in cosmology, eventually produced by the GR and $\Lambda$CDM model. This model has been studied as an alternative to the accelerating expansion because DG can fit the observational SNe-Ia data, and it does not require DE to explain the acceleration of the Universe because it appears naturally from the equations (\cite{universe5020051}). In this work we calculate the scalar TT CMB spectrum and analyze the results and the physical implications. This spectrum is a crucial evidence because it gives us information about the constituents of the Universe and allows us to constraint the DG model. 
In Section \ref{sec:definition}, we show a summary of DG and develop some critical definitions and characteristics of the model. In section \ref{sec:cosmology} we review some important concepts defined in \cite{universe5020051} and also we include new definitions related to the physical densities and thermodynamics in DG. In Section \ref{sec:supernovas}, we show some new results related to SNe-Ia data. These results are slightly different from the work \cite{universe5020051} and they are vital to analyze if the CMB spectrum is in concordance with SNe-Ia. In Section \ref{sec:CMB} we develop some aspects related to the CMB fluctuations and we calculate the scalar TT CMB spectrum assuming a hydrodynamical approach. Finally, we discuss the results and the compatibility between the scalar TT CMB spectrum and the SNe-Ia data.\medskip

\section{Delta Gravity model} \label{sec:definition}

In a previous work, \cite{DeltaGravityL} studied a model of gravitation that is very similar to classical GR but could make sense at the quantum level. In this construction, he considered two different points. The first is that GR is finite on shell at one loop (\cite{tHooft}), then renormalization is not necessary at this level. The second is a type of gauge theories, $\tilde{\delta}$ Gauge Theories (Delta Gauge Theories, \cite{DGT1, DGT2}), which main properties are: (a) New kinds of fields are created, $\tilde{\phi}_I$, from the originals $\phi_I$. (b) The classical equations of motion of $\phi_I$ are satisfied in the full quantum theory.  (c) The model lives at one loop.  (d) The action is obtained by extending the original gauge symmetry of the model, introducing an extra symmetry that we call $\tilde{\delta}$ symmetry since it is formally obtained as the variation of the original symmetry. When we apply this prescription to GR, we obtain DG.\medskip 

We studied the classical effects of DG at the cosmological level. For this, we assume that the Universe is composed of non-relativistic matter (DM and baryonic matter) and radiation (photons and massless particles), which satisfy a fluid-like equation $p = \omega \rho$. Matter dynamics are not considered, except by demanding that the energy-momentum tensor of the matter fluid is covariantly conserved. In \cite{universe5020051} we used the exact solution of the equations, corresponding to the above suppositions, to fit the SNe-Ia data and we obtained an accelerated expansion of the Universe in the model without DE. We have to redefine some important equations and introduce important modifications with respect to previous works.\medskip

These modified theories consist of the application of a variation represented by $\tilde{\delta}$. It has all the properties of a common variation such as:

\begin{eqnarray}
\tilde{\delta}(AB)&=&\tilde{\delta}(A)B+A\tilde{\delta}(B), \nonumber \\
\tilde{\delta}\delta A &=&\delta\tilde{\delta}A, \nonumber \\
\tilde{\delta}(\Phi_{, \mu})&=&(\tilde{\delta}\Phi)_{, \mu},
\end{eqnarray}

where $\delta$ is another variation. The particular characteristic with this variation is that, when we apply it on a field (function, tensor, etc.), it will give new elements that we define as $\tilde{\delta}$ fields, which are an entirely new independent object from the original, $\tilde{\Phi} = \tilde{\delta}(\Phi)$. We use the convention that the new tensor is equal to the $\tilde{\delta}$ transformation of the original tensor when all its indexes are covariant.\medskip

First, we need to apply the $\tilde{\delta}$ prescription to a general action. The extension of the new symmetry is given by:

\begin{equation}
S_0 =\int d^n x \mathcal{L}_0(\phi,\partial_i \phi) \to S = \int d^n x \left(\mathcal{L}_0(\phi,\partial_i \phi)+ \tilde{\delta} \mathcal{L}_0(\phi,\partial_i \phi)\right),
\end{equation}

where $S_0$ is the original action, and $S$ is the extended action in Delta Gauge Theories. GR is based on Einstein-Hilbert action:

\begin{equation}
S_0 =\int d^4 x \mathcal{L}_0(\phi)= \int d^4 x \sqrt{-g} \left(\frac{R}{2\kappa}+L_M \right),
\end{equation}

where $L_M = L_M(\phi_I,\partial_{\mu}\phi_I)$ is the Lagrangian of the matter fields $\phi_I$ and $\kappa = \frac{8 \pi G}{c^4}$. Then, the DG action is given by

\begin{eqnarray}
\label{grav action}
S =S_0 + \tilde{\delta} S_0= \int d^4x \sqrt{-g} \left(\frac{R}{2\kappa} + L_M - \frac{1}{2\kappa}\left(G^{\alpha \beta} - \kappa T^{\alpha \beta}\right)\tilde{g}_{\alpha \beta} + \tilde{L}_M\right),
\end{eqnarray}

where we have used the definition of the new symmetry: $\tilde{\phi} = \tilde{\delta}\phi$ and the metric convention of {\cite{weinberg2008cosmology}}\footnote{In \cite{ag} you can find more about the formalism of the DG action and the new symmetry $\tilde{\delta}$.} \footnote{We emphasize that DG is not a metric model of gravity because massive particles do not move on geodesics. Only massless particles move on geodesics of a linear combination of both tensor fields.} and

\begin{eqnarray}
\tilde{g}_{\mu \nu} = \tilde{\delta} g_{\mu \nu},\\
\label{EM Tensor}
T^{\mu \nu} = \frac{2}{\sqrt{-g}} \frac{\delta\left(\sqrt{-g} L_M\right)}{\delta g_{\mu \nu}}, \\
\label{tilde L matter}
\tilde{L}_M = \tilde{\phi}_I\left(\frac{\delta L_M}{\delta \phi_I}\right) + (\partial_{\mu}\tilde{\phi}_I)\left(\frac{\delta L_M}{\delta (\partial_{\mu}\phi_I)}\right),
\end{eqnarray}

where $\tilde{\phi}_I = \tilde{\delta}\phi_I$ are the $\tilde{\delta}$ matter fields (also called called Delta matter fields). Thus, the equations of motion are:

\begin{eqnarray}
\label{Einst Eq} G^{\mu \nu} &=& \kappa T^{\mu \nu}, \\
\label{tilde Eq} F^{(\mu \nu) (\alpha \beta) \rho
\lambda} D_{\rho} D_{\lambda} \tilde{g}_{\alpha \beta} + \frac{1}{2}g^{\mu \nu}R^{\alpha \beta}\tilde{g}_{\alpha \beta} - \frac{1}{2}\tilde{g}^{\mu \nu}R &=& \kappa\tilde{T}^{\mu \nu},
\end{eqnarray}

with

\begin{eqnarray}
\label{F}
F^{(\mu \nu) (\alpha \beta) \rho \lambda} &=& P^{((\rho
\mu) (\alpha \beta))}g^{\nu \lambda} + P^{((\rho \nu) (\alpha
\beta))}g^{\mu \lambda} - P^{((\mu \nu) (\alpha \beta))}g^{\rho
\lambda} - P^{((\rho \lambda) (\alpha \beta))}g^{\mu \nu} \nonumber, \\
P^{((\alpha \beta)(\mu \nu))} &=& \frac{1}{4}\left(g^{\alpha
\mu}g^{\beta \nu} + g^{\alpha \nu}g^{\beta \mu} - g^{\alpha
\beta}g^{\mu \nu}\right) \nonumber,\\
\tilde{T}^{\mu \nu}&=&\tilde{\delta} T^{\mu \nu}, \nonumber
\end{eqnarray}

where $(\mu \nu)$ denotes that $\mu$ and $\nu$ are in a totally symmetric combination. The DG equations are of second order in derivatives which is needed to preserve causality and the Equation $\eqref{tilde Eq}_{\mu \nu} = \tilde{\delta}\left[\eqref{Einst Eq}_{\mu \nu}\right]$. Also, there are two conservation rules given by:

\begin{eqnarray}
\label{Conserv T}
D_{\nu}T^{\mu \nu} &=& 0 \\
\label{Conserv tilde T}
D_{\nu}\tilde{T}^{\mu \nu} &=& \frac{1}{2}T^{\alpha \beta}D^{\mu}\tilde{g}_{\alpha \beta} - \frac{1}{2}T^{\mu \beta} D_{\beta}\tilde{g}^{\alpha}_{\alpha} + D_{\beta}(\tilde{g}^{\beta}_{\alpha}T^{\alpha \mu})
\end{eqnarray}

It is easy to see that the Equation \eqref{Conserv tilde T} is $\tilde{\delta}\left(D_{\nu}T^{\mu \nu}\right) = 0$.

\subsection{\texorpdfstring{$T^{\mu\nu}$}{Tun} and \texorpdfstring{$\tilde{T}^{\mu\nu}$}{TTun} for a perfect fluid}

In DG, the energy-momentum tensors for a perfect fluid are \footnote{Where $c=1$ is the speed of light.}:

\begin{equation}
T_{\mu\nu}\ =\ p(\rho)g_{\mu\nu}+\left(\rho+p(\rho)\right)U_{\mu}U_{\nu}
\end{equation}

\begin{equation}
\begin{split}
\tilde{T}_{\mu \nu}\ = p(\rho)\tilde{g}_{\mu \nu}+\frac{\partial p}{\partial\rho}(\rho)\tilde{\rho}g_{\mu \nu}+\left(\tilde{\rho}+\frac{\partial p}{\partial\rho}(\rho)\tilde{\rho}\right)U_{\mu}U_{\nu}+\\
\left(\rho+p(\rho)\right)\left(\frac{1}{2}(U_{\nu}U^{\alpha}\tilde{g}_{\mu\alpha}+U_{\mu}U^{\alpha}\tilde{g}_{\nu\alpha})+U_{\mu}^{T}U_{\nu}+U_{\mu}U_{\nu}^{T}\right)
\end{split}
\end{equation}

where $U^\alpha U_\alpha ^T=0$. $p$ is the pressure, $\rho$ is the density and $U^\mu$ is the four-velocity. For more details you can see \cite{DeltaGravityL}.

\subsection{Geodesic equation for massless particles}

In DG, a massless particle behaves according to the following equation:

\begin{equation}
\mathbf{g}_{\mu\nu}\dot{x}^{\mu}\dot{x}^{\nu}=0,
\label{ecmovsinmasa}
\end{equation}

Where the Effective Metric $\mathbf{g}_{\mu\nu}$ is a linear combination given by the two tensors: 

\begin{equation}
\mathbf{g}_{\mu\nu}=g_{\mu\nu}+\tilde{g}_{\mu\nu}
\label{MasslessParticlesGeodesic}
\end{equation}

Thus, the massless particles follow null geodesic, like in the GR theory. We remark that massive particles do not follow geodesics \cite{DeltaGravity}.

\section{Cosmology in Delta Gravity} \label{sec:cosmology}

\subsection{Effective Metric to describe the Universe in a cosmological frame}

The usual metric to describe the Universe in the standard cosmology is the FLRW metric. We assume a flat Universe ($k=0$), then the metric is  given by the Equation \eqref{FLRWmetric}:

\begin{equation}
\mathrm{d}s^2=g_{\mu \nu}\mathrm{d}x^\mu \mathrm{d}x^\nu = -c^2\mathrm{d}t^2 + {a(t)}^2 \left(dx^2+dy^2+dz^2\right),
\label{FLRWmetric}
\end{equation}

where the Scale Factor is called $a(t)$.\\

The objective is to build an Effective Metric for the Universe; then the equations need to explain the photon trajectories, because these particles are what we observe and provide us the information from the observables (such as the SNe-Ia data), showing us the expansion of the Universe. As in the GR frame, we build the metric for the Universe using the massless particle geodesic in DG. We have to include a ``scale factor'' in the space-metric component to explain the expansion of the Universe. This factor must be space-independent because we want to preserve the homogeneity and isotropy for the Universe, then it has to be time-dependent. Therefore, we have to find $\tilde{g}_{\mu \nu}$ from the $g_{\mu \nu}$. We are going to do a change of variable in the Standard Metric tensor, $t\to u$, where  $T(u)= \frac{dt}{du}(u)$:

\[g_{\mu\nu}dx^{\mu}dx^{\nu}=-T^{2}(u)c^{2}du^{2}+a^{2}(u)(dx^{2}+dy^{2}+dz^{2}).\]

Now we add the new dependencies to the temporal and spatial components of the equation, building the most general metric without losing the  homogeneity and isotropy of the Universe:

\[\tilde{g}_{\mu\nu}dx^{\mu}dx^{\nu}=-F_{b}(u)T^{2}(u)c^{2}du^{2}+F_{a}(u)a^{2}(u)(dx^{2}+dy^{2}+dz^{2}), \]

thus, we have to fix a gauge to delete the extra degrees of freedom. Fixing an Harmonic gauge (described in \cite{DeltaGravity}) we obtain:

\[T(u)=T_{0}a^{3}(u),\]

\[F_{b}(u)=3(F_{a}(u)+T_{1}),\]

where $T_{0}$ and $T_{1}$ are gauge constants. Choosing $T_{0}=1$ and $T_{1}=0$ the gauge is fully fixed.
Finally, we go back to the Effective Metric described by the Equation \eqref{MasslessParticlesGeodesic} to substitute the fixed gauges. This defines the Effective Metric for the Universe in DG:

\begin{equation}
\mathbf{g}_{\mu \nu}=g_{\mu\nu}+ \tilde{g}_{\mu\nu}=-\left(1+3F(t)\right)c^2 dt^2+a^2(t) \left(1+F(t))\right)\left(dx^2+dy^2+dz^2\right),
\label{EffectiveMetric}
\end{equation}

where the proper time is determined by the original tensor $g_{\mu\nu}$ (\cite{Alfaro:2010kv}).

\subsection{Delta Gravity equations of motion}

\label{explanation_DG_densities}

To apply this theory to cosmology, we assume that the Universe has two components: matter and radiation. With the new symmetry, two kinds of new components appear: Delta matter and Delta radiation.\medskip

To calculate the equations that govern the Universe, we assume that $g_{\mu \nu}$ is expressed by the Equation \eqref{FLRWmetric} and we calculate the First Field Equation given by the Equation \eqref{Einst Eq}:

\begin{equation}
\left(\frac{\dot{a}(t)}{a(t)}\right)^2=\frac{\kappa c^4}{3} \left(\rho_{r}(t)+\rho_{m}(t)\right).
\label{EcDeltaFriedmann1}
\end{equation}

If we solve the Equation \eqref{EcDeltaFriedmann1}, we obtain the following expression:

\begin{equation}
\dot{\rho}_{i}(t)\ =\ -\frac{3\dot{a}(t)}{a(t)}(\rho_{i}(t)+p_{i}(t)).
\end{equation}

Considering an equation of state, it is possible to relate  $\rho$ and $p$ for each component $i$. Assuming that there are only matter (baryonic, and  dark matter) and radiation (photons and other massless particles), we have (same as GR at this point) for matter:

\[p_m(a)=0,\]

and for radiation:

\[p_r(a)=\frac{1}{3}\rho_r(a).\]

With these equations we can solve the Equation \eqref{EcDeltaFriedmann1} expressing $t(a)$. Summarizing, we have:

\begin{eqnarray}
\rho(a)=\rho_m(a)+\rho_r(a),\label{177}\\
p_r(a)=\frac{1}{3}\rho_r(a),\label{178}\\
t(Y)=\frac{2\sqrt{C}}{3H_0\sqrt{\Omega_{r,0}}}\left(\sqrt{Y+C}(Y-2C)+2C^{3/2}\right),\label{CosmologicalTimeDeltaGravity}\\
Y(t)=\frac{a(t)}{a_0}, \label{180}\\
a_0\equiv a(t=t_0)\equiv 1,\\
\Omega_{r,0}\equiv \frac{\rho_{r,0}}{\rho_{c,0}},\\
\Omega_{m,0}\equiv \frac{\rho_{m,0}}{\rho_{c,0}},\\
\rho_{c,0}\equiv  {\frac {3H_0^{2}}{8\pi G}}\label{CriticalDensity},\\
\Omega_{r,0}+\Omega_{m,0}\equiv 1,
\end{eqnarray}

where $t_0$ is the age of the Universe (today).
We emphasize that $t$ is the Cosmic Time, $a_0$ is the Scale Factor today, $C\equiv \frac{\Omega_{r,0}}{\Omega_{m,0}}$, where  $\Omega_{r,0}$ and $\Omega_{m,0}$ are the density energies normalized by the critical density today, defined as the same as the standard cosmology. Furthermore, we have imposed that Universe must be flat ($k=0$), so we require that $\Omega_{r,0}+\Omega_{m,0}\equiv 1$. Note that $\rho_i$ is not a physical density. They are only density parameters that are related to physical densities \footnote{They are not energy per volume.}. We are going to discuss this aspect in the next pages.\\

Using the second continuity Equation \eqref{Conserv tilde T}, where $\tilde{T}_{\mu \nu}$ is a new energy-momentum tensor, we define two new densities called $\tilde{\rho}_m$ (Delta matter density) and $\tilde{\rho}_r$ (Delta radiation density). They are associated with this new tensor. When we solve this equation, we find

\begin{eqnarray}
\tilde{\rho}_m(Y)= \frac{C_1-\frac{3}{2}\rho_{m,0}F(Y)}{Y^3} \label{183},\\
\tilde{\rho}_r(Y)=\frac{C_2-2\rho_{r,0}F(Y)}{Y^4} \label{184}
\end{eqnarray}

where $C_1$ and $C_2$ are integration constants.
It is crucial to clarify that $\tilde{\rho}_m$ and $\tilde{\rho}_r$ depend on the Normalized Scale Factor $Y$. We can note that both energy density parameters (remember that these parameters are not real physical densities. But they are related to the physical densities) have terms that behave like the standard cosmology densities $\sim\frac{1}{Y^3}$ and $\sim\frac{1}{Y^4}$ that also are preserved in DG:

\begin{equation}
\rho_r(Y) = \frac{\rho_{r,0}}{Y^4}
\end{equation}

\begin{equation}
\rho_r(Y) = \frac{\rho_{m,0}}{Y^3}
\end{equation}

If we preserve $C_1 \neq 0$ and $C_2\neq 0$, we have equations that are considering two kinds of dependence: $\sim\frac{1}{Y^3} +\frac{F(Y)}{Y^3}$ and $\sim\frac{1}{Y^4} +\frac{F(Y)}{Y^4}$. This consideration implies that the total energy density (proportional to the real physical densities) considers the standard energy density and the new dependence given by DG, in other words, this is equivalent to consider that $\tilde{\rho}_r$ is the standard density radiation $\rho_r$ plus the new DG dependence. We only want to consider the new dependence in the $\tilde{\rho}_r$ term without the standard radiation contribution. This same reasoning is valid for the density of matter. Thus, defining $C_1=C_2=0$, we obtain the following equations:

\begin{eqnarray}
\tilde{\rho}_m(Y)= -\frac{3\rho_{m,0}}{2}\frac{F(Y)}{Y^3} \label{DeltaMatterDensity},\\
\tilde{\rho}_r(Y)=-2\rho_{r,0}\frac{F(Y)}{Y^4}. \label{DeltaRadiationDensity}
\end{eqnarray}

There is another reason to define $C_1$ and $C_2$ equal to $0$. When $Y\ll C$, the Effective Scale Factor $Y_{DG}$ (defined in Equations \eqref{YtildeDef} and \eqref{RtildeFa}) represents the evolution of the Universe at the beginning. We know that an accelerated expansion appears at late times, then the non-relativistic matter and radiation must drive the expansion at early times, this means $Y_{DG} = 1 + O(Y )$. We fix $C_1 = 0$ and $C_2 = 0$ to guarantee that the behavior of expansion seems like GR at early times. The full development of this idea can be found in \cite{Alfaro_2012,Alfaro2013}.\\

Using the Equation \eqref{tilde Eq} with the solutions from the Equations \eqref{DeltaMatterDensity} and \eqref{DeltaRadiationDensity} we found (and redefining with respect to $Y$):

\begin{equation}
F(Y)=-\frac{L_2}{3}Y\sqrt{Y+C},
\label{Fa-def}
\end{equation}

where $L_2$ is an arbitrary constant.

\subsection{Relation between the Effective Scale Factor \texorpdfstring{${Y_{DG}}$}{Lg} and the Normalized Scale Factor \texorpdfstring{$Y$}{Lg}}

The Effective Metric for the Universe is given by the Equation \eqref{EffectiveMetric}. From this expression, it is possible to define the DG Scale Factor as follows:

\begin{equation}
{a_{DG}}(t)=Y(t)\sqrt{\frac{1-\frac{L_2}{3}Y\sqrt{Y+C}}{1-L_{2}Y\sqrt{Y+C}}}.
\label{RtildeFa}
\end{equation}

Furthermore, we define the Effective Scale Factor as $a_{DG}$ normalized by itself:

\begin{equation}
{Y_{DG}}(t)\equiv\frac{{a_{DG}}(t)}{{a_{DG}}(t_0)}.
\label{YtildeDef}
\end{equation}

With the new definition of $L_2$, the Delta densities are given by:

\begin{eqnarray}
\tilde{\rho}_m(Y)= \left(\frac{L_2}{2}\right)\rho_{m,0}\frac{\sqrt{Y+C}}{Y^2}, \label{DeltaMatterDensityFinal}\\
\tilde{\rho}_r(Y)=\left(\frac{2L_2}{3}\right)\rho_{r,0}\frac{\sqrt{Y+C}}{Y^3}. \label{DeltaRadiationDensityFinal}
\end{eqnarray}

If we know $C$ and $L_2$, it is possible to calculate the Delta densities $\tilde{\rho}_m$ and $\tilde{\rho}_r$ as function of the common densities.
We emphasize that the denominator in the Equation \eqref{YtildeDef} is equal to zero when $1=L_2 Y\sqrt{Y+C}$. Taking into account that $C=\Omega_{r,0}/\Omega_{m,0}\ll 1$, if $Y=1$ (current time) then the denominator goes to $0$ when $L_2\approx 1$. Furthermore, we have imposed that $\tilde{\rho}_m>0$ and $\tilde{\rho}_r>0$, then $L_2$ must be greater than $0$. Then the valid range for $L_2$ is approximately $0\leq L_2 < 1$.\\
Regarding the $C$ value, it must be a small positive number because the radiation is not dominant compared to matter. Then, we can analyze cases close to the standardly accepted value for $\Omega_{r,0}/\Omega_{m,0}\sim 10^{-4}$ (we have assumed GR values to estimate an order of magnitude).

\subsection{Useful equations for cosmology}

Here we present useful equations to fit the SNe-Ia data and to obtain the cosmological parameters.

\subsubsection{Redshift dependence}

DG preserves the relation between the cosmological redshift and the Effective Scale Factor. The reason is straightforward: it is the same as in GR, but changing the Scale Factor $a(t)\to {a_{DG}}(t)$ in the GR metric $g_{\mu\nu}dx^\mu dx^\nu \to \mathbf{g}_{\mu\nu}dx^\mu dx^\nu$ \cite{DeltaGravity}. Thus, the dependence is given by:

\begin{equation}
{Y_{DG}}(t)=\frac{1}{1+z}.
\label{Ytilderedshift}
\end{equation}

It is important to consider that the current time is given by $t_0\to Y(t_0)\to {Y_{DG}}(Y=1)=1$.

\subsubsection{Luminosity distance}

\label{Sec_AngularDistance_DG} 

The proof is the same as GR, because the main idea is based on the light traveling through a null geodesic described by the Effective Metric given by the Equation \eqref{EffectiveMetric} in DG. Then, the equation that describes the luminosity distance for DG is the same as GR, but changing the Scale Factor $a(t)$ by the ${a_{DG}}(t)$, because ${a_{DG}}(t)$ is the factor that is describing the observable expansion (or scaling) of the Universe \cite{universe5020051}.

We remark that the relation between the luminosity distance $d_L^{DG}$ and angular distance $d_A^{DG}$ in DG is the same as in GR (\cite{doi:10.1080/14786443309462220}). This relation is a direct consequence of the structure of the metric. This relation is given by the Equation \eqref{dLdARelation},

\begin{equation}
{\displaystyle d_{L}^{DG}=(1+z)^{2}d_{A}^{DG}}.
\label{dLdARelation}
\end{equation}

The luminosity distance was calculated in \cite{universe5020051} and is given by

\begin{equation}
d_L^{DG}(z,L_2,C,h^2\Omega_{m,0})=c\frac{(1+z)}{100\sqrt{h^2\Omega_{m,0}}} \int_{Y(t_{1})}^{1}  \frac{Y}{\sqrt{Y+C}}\frac{dY}{{Y_{DG}}(t)}.
\label{LuminosityDistanceDG}
\end{equation}

where $Y=1$ denotes today. To solve $Y(t_1)$ at a given redshift $z$, we need to solve the Equations \eqref{YtildeDef} and \eqref{Ytilderedshift} numerically. Furthermore, the integrand contains the Effective Scale Factor ${Y_{DG}}(t)$ that can be expressed in function of $Y$ through the Equation \eqref{YtildeDef}. Do not confuse $c$ (speed of light) with $C$.
If the integration assumes $Y\gg C$ (a good approximation for SNe-Ia, because we are integrating in late times), this equation can be approximated to:

\begin{equation}
d_L^{DG}(z,L_2,h) \approx c\frac{(1+z)}{100h} \int_{Y(t_{1})}^{1}  \frac{\sqrt{Y}}{{Y_{DG}}(t)} dY,
\label{LuminosityDistanceDGapprox}
\end{equation}

where $d_L^{DG}$ is independent of $C$. Also, if $C \to 0$, then $\Omega_{m,0} = 1/(1+C) \to 1$. In this scenario, the only two free parameters are $h$ and $L_2$.

We underline that it is impossible to know the $C$ using the SNe-Ia data, but we can constraint this value with the CMB. In DG, $C$ is a constant that is related with the physical densities, but it does not represent a ratio between physical densities.

\subsection{Distance modulus}

This relation is fundamental because it lets us calculate the dependence between the apparent magnitude and the distance to the object. It is essential to consider that we need to know the value of the absolute magnitude $ M $ to avoid degeneration.

\begin{equation}
{\displaystyle \mu =m-M=5\log _{10}\left({\frac {d_L^{DG/GR}}{10\,\mathrm {pc} }}\right)}
\label{DistanceModulus}
\end{equation}

\subsection{Normalized Effective Scale Factor}

In DG, the ``size'' of the Universe is given by ${Y_{DG}}(t)$, then every cosmological parameter that in the GR theory was built up from the standard scale factor $a(t)$, in DG will be built from ${Y_{DG}}(t)$.

\subsection{Hubble Parameter} \label{hubble}

The Hubble parameter (and also, the Hubble Constant) is defined in GR cosmology as:

\begin{equation}
H(t)=\frac{\dot{a}(t)}{a(t)}.
\end{equation}

Thus, in DG we define the Hubble Parameter as follows:

\begin{equation}
H^{DG}(t)\equiv\frac{{\dot{a}_{DG}}(t)}{{a_{DG}}(t)}.
\end{equation}

The Hubble Constant is the Hubble parameter $H^{DG}(t)$ evaluated today, in other words, when $Y=1$. Therefore, the Hubble parameter is given by

\begin{equation}
H^{DG}(t)=\frac{\frac{d {a_{DG}}}{dY}\left(\frac{dt}{dY}\right)^{-1}}{{a_{DG}}}.
\label{HDG}
\end{equation}

Observe that all the DG parameters are written as a function of $Y$.

\subsection{Deceleration parameter}

In the standard cosmology the Deceleration parameter is given by:

\begin{equation}
q(t)=-\frac{\ddot{a}a}{\dot{a}^2}.
\label{GRqparameter}
\end{equation}

Thus, in DG we define the Deceleration parameter as follows:

\begin{equation}
q^{DG}(t)=-\frac{{\ddot{a}_{DG}}{a_{DG}}}{{\dot{a}_{DG}}^2}.
\end{equation}

Thus,

\begin{equation}
q^{DG}(t)=-\frac{\frac{d }{dY}\left(\frac{d {a_{DG}}}{dY}\left(\frac{dt}{dY}\right)^{-1}\right)\left(\frac{dt}{dY}\right)^{-1}{a_{DG}}}{\left(\frac{d {a_{DG}}}{dY}\left(\frac{dt}{dY}\right)^{-1}\right)^2}
\label{DGqparameter}
\end{equation}

\subsection{Dependence between redshift and Cosmic Time} \label{tiemporedshift}

All the equations are parametrized as a function of $Y$, so we need to use the Equations \eqref{CosmologicalTimeDeltaGravity}, \eqref{YtildeDef} and \eqref{Ytilderedshift} to relate redshift and Cosmic Time. In the matter-dominated Universe until today, $C\sim 10^{-4}\ll Y$, then

\begin{equation}
t(Y)=\frac{2}{3H_0}Y^{3/2}
\label{t_de_Y}
\end{equation}

This $H_0$ constant is not the Hubble constant. In DG, this is an arbitrary constant that can be obtained from the SNe-Ia fit. it is different from $H^{DG}_0$, the physical and observable constant. The age of the Universe can be easily calculated from the Equation \eqref{t_de_Y} where $L_2$ does not play role in the time evolution.

\subsection{non-physical densities}

\subsection{Non-physical Densities of Common Components: \texorpdfstring{$\Omega_{m,0}$}{Lg} and \texorpdfstring{$\Omega_{r,0}$}{Lg}}

We have imposed that $\Omega_{m,0}+\Omega_{r,0}=1$ and $C=\frac{\Omega_{r,0}}{\Omega_{m,0}}$, then

\begin{equation}
\Omega_{r,0}=\frac{C}{1+C} ;\qquad \Omega_{m,0}=\frac{1}{1+C}.
\end{equation}

It is vital to consider that this equation only expresses a relation, or a proportion, between the non-physical energy density for Common matter and Common radiation densities, and does not express a real percentage of composition of the Universe because in DG we also have Delta matter and Delta radiation.\\

This condition is imposed when we assumed that $T^{\mu\nu}$ only expresses a standard composition, and when we assumed that the DE does not exist either at the level of Action or Field Equations.

\subsection{physical densities in DG: a thermodynamic approach}

This definition is essential to define any physical interaction that is related with the physical parameters, for example, damping associated to fluids or collision probabilities between particles. Thus, this is essential to fit the CMB spectrum.

The physical element of volume in DG is $dV = a_{DG}^3 dx dy dz$ (given by the effective metric), which is described by the DG Scale Factor $a_{DG}$. Then, the density of any kind of matter in terms of energy per volume is

\begin{equation}
    \rho_{DG}=\frac{U}{c^2V},
\end{equation}

where $U$ is the internal energy, and $V$ is the volume and $\rho_{DG}$ is the physical density.\footnote{Do not confuse: the common density $\rho$ in the background, the Delta density $\tilde{\rho}$ and $\rho_{DG}$ the physical density. The latter is the observational density.} Therefore, if we apply the first law of thermodynamics and assume that the evolution of the Universe is adiabatic 
as in GR\footnote{Where the process is isentropic.} (\cite{padmanabhan_2002}),

\begin{equation} \label{energyDG}
\dot{\rho}_{DG}=-3H_{DG}\left(\rho_{DG}+\frac{P_{DG}}{c^2}\right).
\end{equation}

In the standard cosmology, the equations of state are written as $P=\omega \rho$, thus in DG we assume an equation as $P_{DG}=\omega \rho_{DG}$ and replace it in Equation \eqref{energyDG}, then we obtain

\begin{equation}\label{DG density solution}
    \rho_{DG} a_{DG}^{3(1+\omega)}=\rho_{DG,0} a_{DG,0}^{3(1+\omega)},
\end{equation}

where $\rho_{DG,0}$ is the density today. We can relate the physical and the background densities by the ratio between them

\begin{equation}\label{relation densities phys vs GR}
    \frac{\rho_{DG}}{\rho}\left(\sqrt{\frac{1+F(t)}{1+3F(t)}}\right)^{3(1+\omega)}= \text{constant}(\omega).
\end{equation}

The standard cosmological perturbations are defined as

\begin{equation}
    \delta_{\alpha}=\frac{\delta \rho_{\alpha}}{\bar{\rho}_{\alpha}+\bar{p}_{\alpha}},
\end{equation}

where $\alpha = \gamma$, $\nu$, $B$ or $D$ (photons, neutrinos, baryons and dark matter, respectively). In the early Universe, when $Y \sim 10^{-3}$ (near the Last Scattering surface), the $F$ factor tends to $1$. This aspect is vital for the development of the perturbative equations, because at that moment the physical densities were proportional to the standard densities, and  by definition, the physical perturbations are equal to the standard perturbation:

\begin{equation}
    \delta_{phys\;\alpha}(t)=\delta_{\alpha}(t).
\end{equation}

This approximation is accurate and it is valid from the beginning of the Universe ($z\rightarrow \infty$)  to $z\sim 10$.

\subsection{The shape of the black body spectrum}

We want to preserve the shape of the CMB black body spectrum because it is an observable, which is described by

\begin{equation}
    n_T(\nu)d\nu=\frac{8\pi \nu^2d\nu}{e^{\frac{h\nu}{k_BT}}-1}.
\end{equation}

After the Last Scattering surface, the photons traveled decoupled with baryons, then the spectrum changes its frequency as $\nu=\nu_{ls}a_{DG}(t_{ls})/a_{DG}$, but the volume also changes as $V=V_{ls}a_{DG}^3/a_{DG}^3(t_{ls})$, then, the conservation of the number of photons $dN= n_T(\nu) d\nu dV$ implies that 

\begin{eqnarray}\label{Temperature-redhift}
T=\frac{T_0}{Y_{DG}},
\end{eqnarray}

where $T_0$ is the CMB temperature. In other words, the temperature of the Universe evolves with the Effective Scale Factor described by $Y_{DG}$ and not $a$.\medskip

All these definitions and interpretations are essential to describe the CMB physics. The preservation of the relation given by \eqref{Temperature-redhift} is important because the deviation of the $T$ with $z$ has been previously studied in \cite{TvszLima} as an arbitrary dependence in the $T$, where the results found by \cite{TvszdeMartino,TvszAvgoustidis_2012} indicated that $T = T_0 (1+z)$ is right.

We are interested in the viability of DG as a real alternative cosmology theory that could explain the accelerating Universe without $ \Lambda $. The first Section shows the SNe-Ia data and the equations, the Section 2 shows the results and the last Section contains the analysis and the conclusions. This chapter is similar to the previous one, but the meaning of some parameters and their numerical values change. This change is relevant to be able to explain the CMB later.

\section{SNe-Ia and the accelerated expansion} \label{sec:supernovas}

\label{sec:FittingTheSNData}

\subsection{SNe-Ia data}

The SNe-Ia are very useful in cosmology  because they can be used as standard candles allowing to fit a cosmological model (\cite{Riess1998}). The main characteristic of the SNe-Ia that makes them so useful is that they have a very standardized absolute magnitude close to $-19$ (\cite{Riess2016, Mabs-19.05,Mabs-19.05BetoulePeroEnLibro,Mabs-19.25,Mabs-19.26}).

To analyze the expansion of the Universe, we used 1048 SNe from the type Ia supernovae catalog from \cite{Scolnic2018}. We only need to know the distance modulus $\mu$ and the redshift $z$ for every SN to fit the DG model using the luminosity distance given in the Equation \eqref{DistanceModulus}. We assume a scenario with  $M$ fixed and a flat Universe where the radiation is negligible ($C=0$ for DG, and $\Omega_{r,0} = 0$ for GR), and fit the DG model to find $L_2$ and $h$ while in GR we find $\omega_{m,0}$ and $h$. The $M$ value was calculated using 210 SNe-Ia from \cite{Riess2016} and corresponds to the absolute magnitude which is independent of the model. \footnote{This value is independent of the cosmological model because it was calculated building the distance ladder from local Cepheids measured by parallax and using them to calibrate the distance to Cepheids hosted in nearest galaxies (by period-luminosity relations) that are also SN-Ia host (\cite{Riess2016}) calculated the $M$ and the $H_0$ local value, and they did not use any particular cosmological model.}\medskip

Summarizing, in DG we fit $L_2$ and $h$ while in GR we fit $\Omega_{m,0} = 1 - \omega_{\Lambda}$ and $h$. We emphasize that $h$ in DG model is not the Hubble Constant ($H^{DG}_0$ can be calculated with the Equation \eqref{HDG}), but in the GR case $h$ is the Hubble Constant $H_0$. Both models have two degrees of freedom and for both cases we used Least Squares Method.

\subsection{GR fit}

 In the GR case, the $h$ and $M$ parameters are degenerated. We fix $M$ because it is an independent value obtained from a local measurement and allow us to avoid the degeneration. We include the GR case to compare it with the DG model.\medskip

\begin{figure}[h!]
\centering
\includegraphics[width=16cm,keepaspectratio]{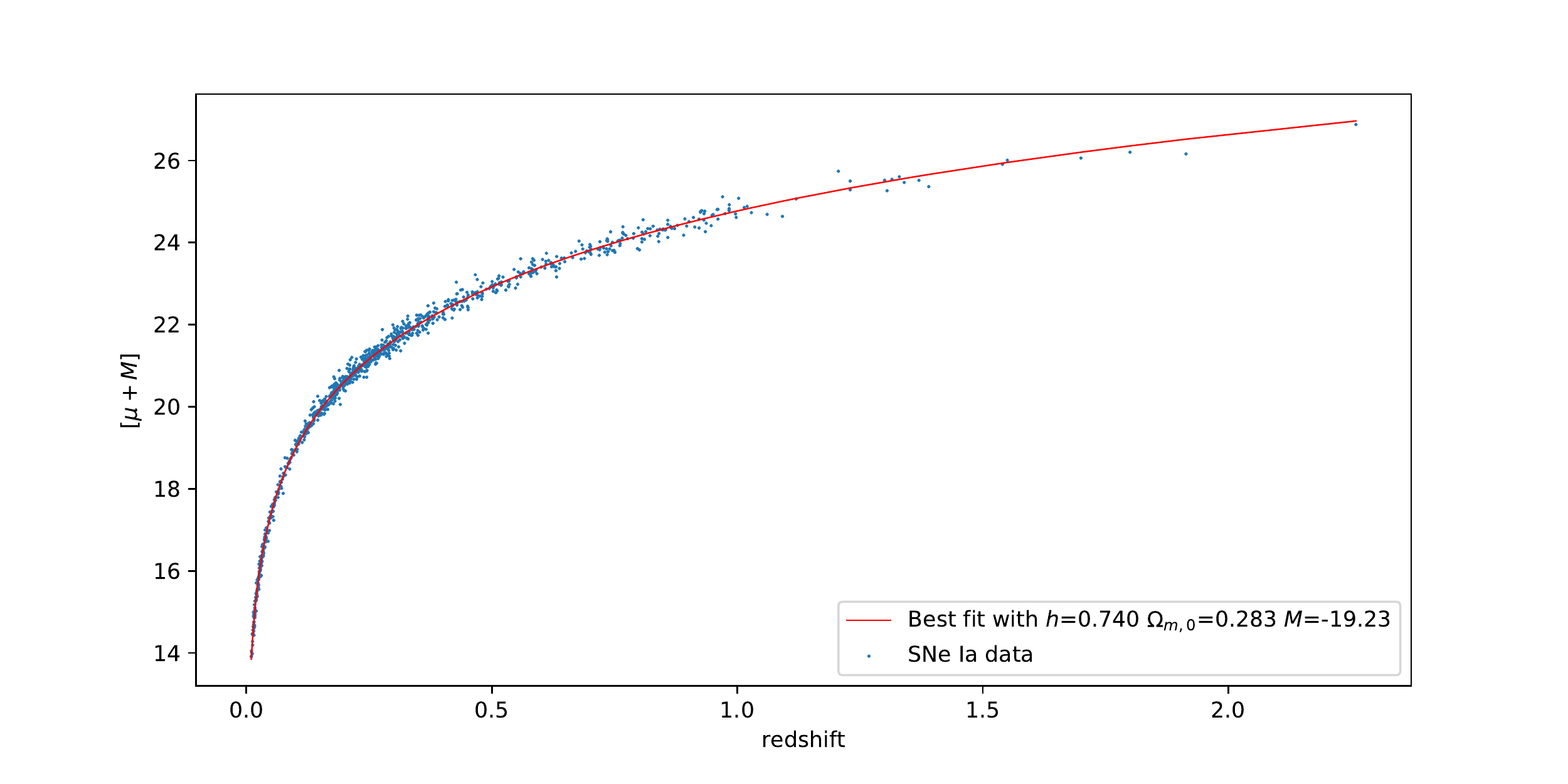}
\caption{The fitted curve for the GR model assumes $M = -19.23$.}
\label{Fig:GRFit}
\end{figure}

The fitted parameters for the GR case are shown in Table \ref{Tab:GR1}.

\begin{table}
\centering
\begin{tabular}{|c|c|c|c|}
\hline
Parameter & Value & Standard Error & Relative Error \\ \hline
$\Omega_{m,0}$        & 0.28 & 0.01    & 4.20\%         \\ \hline
$h$ & 0.740 & 0.002        & 0.33\%         \\ \hline
\end{tabular}
\caption{Fitted values for GR model.}
\label{Tab:GR1}
\end{table}

\subsection{DG fit}

\begin{figure}
\centering
\includegraphics[width=16cm,keepaspectratio]{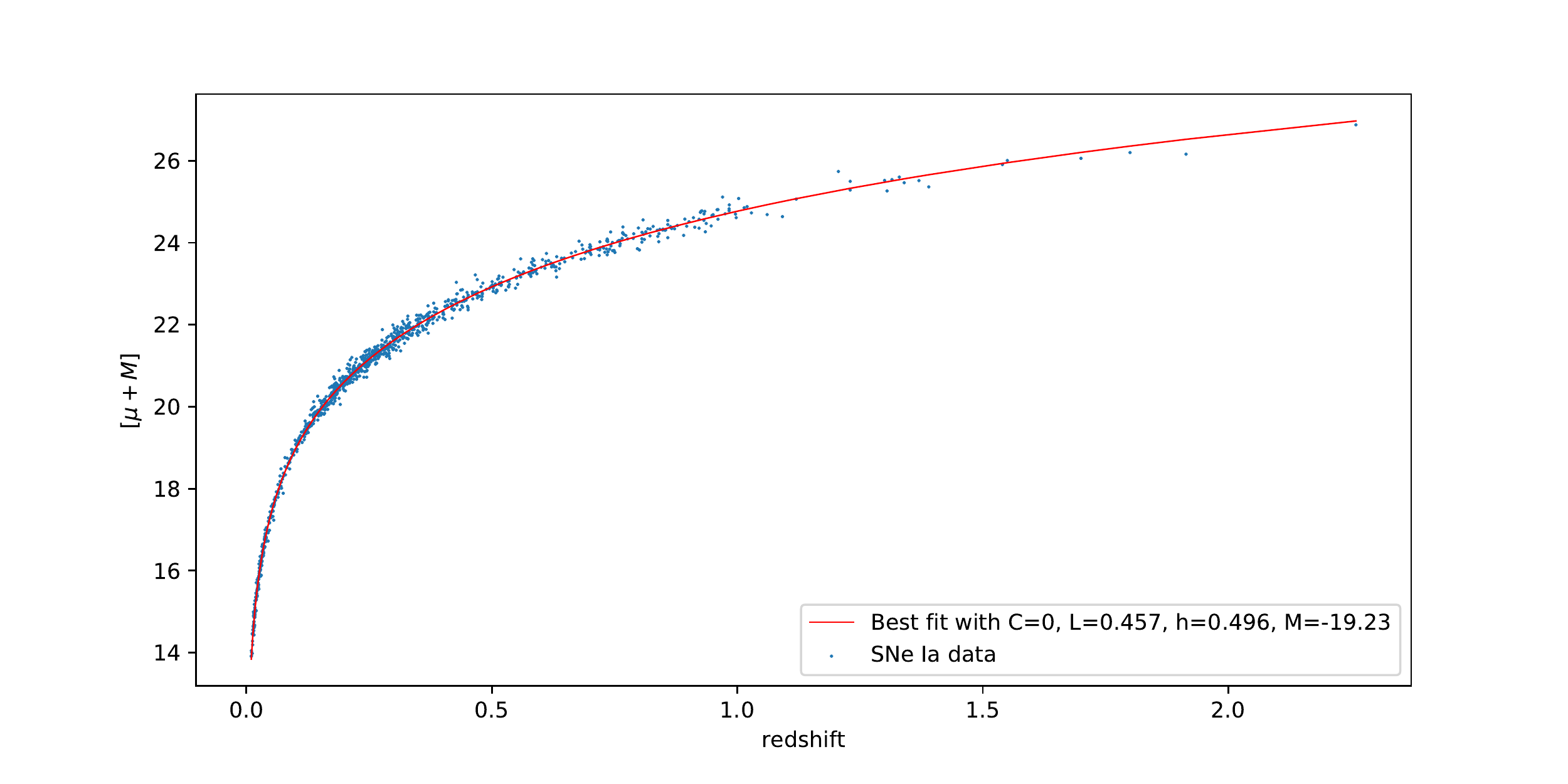}
\caption{The fitted curve for the DG model assumes $C = 0$ and $M = -19.23$.}
\label{Fig:DGFit1}
\end{figure}

We present two figures associated to the DG model. The Figure \ref{Fig:DGFit1} assumes $C = 0$ and  describes very well the SNe-Ia data; DG and GR are indistinguishable in describing the SNe-Ia data. The fitted DG parameters $h$ and $L_2$ are shown in Table \ref{Tab:DG1}. 

\begin{table}
\centering
\begin{tabular}{|c|c|c|c|}
\hline
Parameter & Value & Standard Error & Relative Error \\ \hline
$L_2$         & 0.457 & 0.007          & 1.57\%         \\ \hline
$h$        & 0.496 & 0.004          & 0.77\%         \\ \hline
\end{tabular}
\caption{Fitted values for DG model.}
\label{Tab:DG1}
\end{table}

\begin{figure}
\centering
\includegraphics[width=16cm,keepaspectratio]{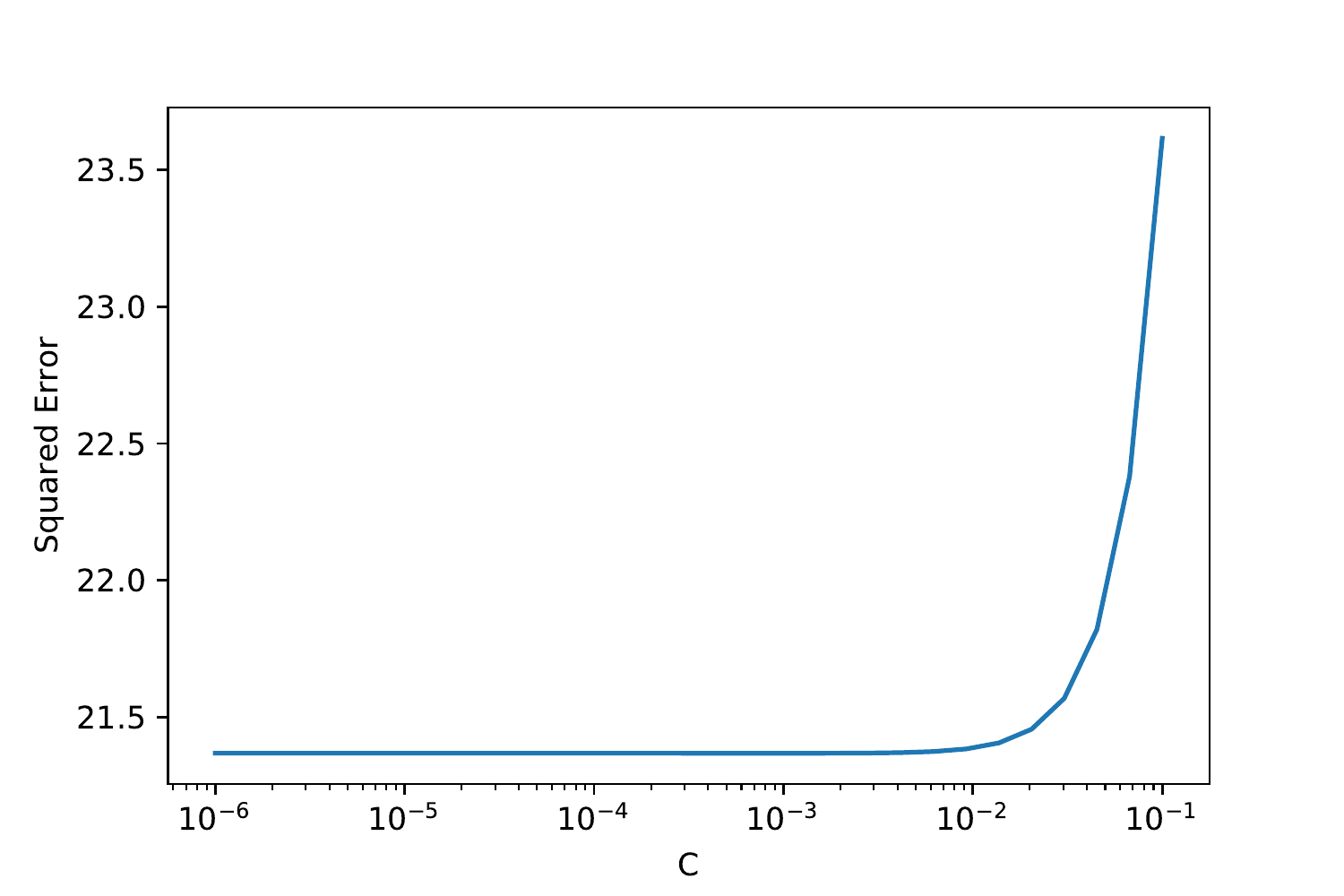}
\caption{Squared errors for DG model assuming a fitted model with $C = 0$,$L_2 = 0.457$ and $M = -19.23$.}
\label{Fig:DGSquaredErrors}
\end{figure}

It is important to analyze the influence of $C \neq 0$ in the approximation that we used. Thus, we show the squared error associated with different $C$ values in the Figure \ref{Fig:DGSquaredErrors}.

The results from SNe-Ia analysis indicate that DG explains the accelerating expansion of the Universe without including $\Lambda$ or anything like ``Dark Energy''. The acceleration is naturally produced in DG, caused by a coefficient named $L_2$, which appears when we solve the differential equations that describe the cosmology. There are two crucial differences between these results and \cite{universe5020051}. Now, the fit assumes $C = 0$ and a physical density different from the background definition, in other words, specifically, $C$ is not a ratio between physical densities. This will be vital in the results of the CMB spectrum.\medskip

An important result from the fitted curves is the independence between the curve fitting and $ C $ value in a wide range of $0 \leq C \ll 10^{-2}$ . If $C$ is about $10^{-4}$ it is impossible to distinguish a curve with $C=0$ or with $C\sim 10^{-4}$. This indistinguishable is crucial because the range of $ C $ allows us to fit the CMB without changing the SNe-Ia fit (if $ C $ is small).

\subsection{Local expansion}

The luminosity distance given by the Equation \eqref{LuminosityDistanceDG} can be simplified assuming $C= 0$ and using the relation between the DG Scale Factor and redshift given by the Equation  \eqref{Ytilderedshift}
we obtain an expression around $z = 0$ up to second order in redshift given by

\begin{equation}
d_L^{DG}(z,L_2,C)\approx \frac{c}{H_0^{DG}} \left( z     +   \frac{1}{2}(1-q_0) z^2\right).
\end{equation}

where at first order, the local expansion is exactly as the $\Lambda$CDM model:

\begin{equation}
    m = 5\log{\frac{cz}{ H_0^{DG}}} + M + 25.
    \label{DG_local_fit_SN}
\end{equation}

This expression is in concordance with the definition given in the Equation \eqref{HDG}.

\cite{Riess2018} found values for $M$ and $H_0$ that are independent of any assumptions (only depends on the $d_L$ definition, where they assumed a flat Universe) and that are not degenerate. Therefore, the local analysis for DG is valid, where the Hubble Constant measured in this context is $H_0^{DG}$ and not $H_0$. Also, note that $H_0$ is very different from $H_0^{DG}$, which is not a problem in DG. Until here, we are trying to conciliate local and high redshift measurements of SNe-Ia data. If any of these observations or data are wrong, all the analyses presented here must be revisited because it depends on both observations.

\subsection{\texorpdfstring{$H^{DG}$}{HDG} and \texorpdfstring{$q^{DG}$}{qDG}}

With the DG fitted parameters we can find $H(t)$ and $H_0$. For GR, $H_0^{GR}$ is easily obtained from the $h^2$ fitted ($H_0=100h$) and $H^{GR}(t)$ can be obtained using the first Friedmann equation

\begin{equation}
H^2 = \left(\frac{\dot{a}}{a}\right)^2 = \frac{8 \pi G}{3}\left(\frac{\rho_{m,0}}{a^3}+\frac{\rho_{r,0}}{a^4}+\rho_{\Lambda,0}\right)
\label{FirstFriedmannEquation}
\end{equation}

Considering that $\Omega_{m,0}+\Omega_{r,0}+\Omega_{\Lambda, 0}=1$, $\Omega_{r,0}\approx 0$ and $\rho_{c,0}=\frac{3 H_0^2}{8\pi G}$, where $\Omega_{i,0}=\frac{\rho_{i,0}}{\rho _{c,0}}$ for every $i$ component in the Universe, we obtain

\begin{equation}
H^2  = H_0^2\left(\frac{\Omega_{m,0}}{a^3}+(1-\Omega_{m,0})\right)
\label{FirstFriedmannEquation2}
\end{equation}

With the Equation \eqref{FirstFriedmannEquation2}, we obtain $H^{GR}(t)$ and using the Equation \eqref{HDG} we obtain $H^{DG}(t)$. To evaluate the Hubble constant we evaluate $H^{GR}$ at $a=1$ for GR and $H^{DG}$ at $Y_{DG}=1$ for DG. The values that we are using in this section are not local, they were obtained using all the SNe-Ia data. Therefore, this GR fit does not imply that $H^{GR}_0$ must be equal to the result obtained by \cite{Riess2016}, however they are similar.\medskip

The $H^{DG}_0$ can be approximated assuming $C = 0$. This estimation is very precise\footnote{This equation is straightforward from the definition of \eqref{HDG}.}:

\begin{equation}
    H^{DG}_0 \approx 50 h \frac{(-6+11 L_2-7 L_2^2+2 L_2^3)}{(-3+L_2)(-1+L_2)^2}
    \label{HDGapprox}
\end{equation}

We present the results from both models, and we compare these values with measurements in the Table  \ref{tab:H0compared}. 
\begin{longtable}{lll}
Model     & $H_0$ ( km/(s Mpc) )  & Error ( km/(s Mpc) )\\
\hline
Planck 2015 \cite{Planck2015} & 67.74  & 0.46  \\
Planck 2018 \cite{Planck2018} & 67.4 & 0.5\\
Riess 2016 \cite{Riess2016} \footnote{First local determination of the Hubble Constant: ``A 2.4\% Determination of the Local Value of the Hubble Constant"} & 73.24 & 1.74\\
Riess 2018\footnote{The calibration was made including the new MW parallaxes from \textit{HST} and \textit{Gaia}.} \cite{Riess2018} & 73.52 & 1.62  \\
Riess 2019\footnote{Precision HST photometry of Cepheids in the Large Magellanic Cloud (LMC) reduce the uncertainty in the distance to the LMC from 2.5\% to 1.3\%} \cite{Riess2018} & 74.03 & 1.42\\
GR & 74.0 & 0.2  \\
DG & 74.3 & 1.3  \\
DG approx & 74.2 & - \\
DG local & 73.5 & 0.4\\
\caption{$H_0$ values found by Least Squares Method with SNe-Ia data. Furthermore, we tabulate 
Planck satellite's data \cite{Planck2015} and  \cite{Planck2018}, and Riess et al. \cite{Riess2018} $H_0$ values. $GR$ and $DG$ are the $H_0$ values obtained in Section \ref{sec:FittingTheSNData} using all the SNe-Ia data. $DG approx$ was calculated from the Equation \eqref{HDGapprox} and $DG local$ was obtained fitting local SNe-Ia using the Equation \eqref{DG_local_fit_SN}.}
\label{tab:H0compared}
\end{longtable}

\begin{figure}
\centering
\includegraphics[width=16cm,keepaspectratio]{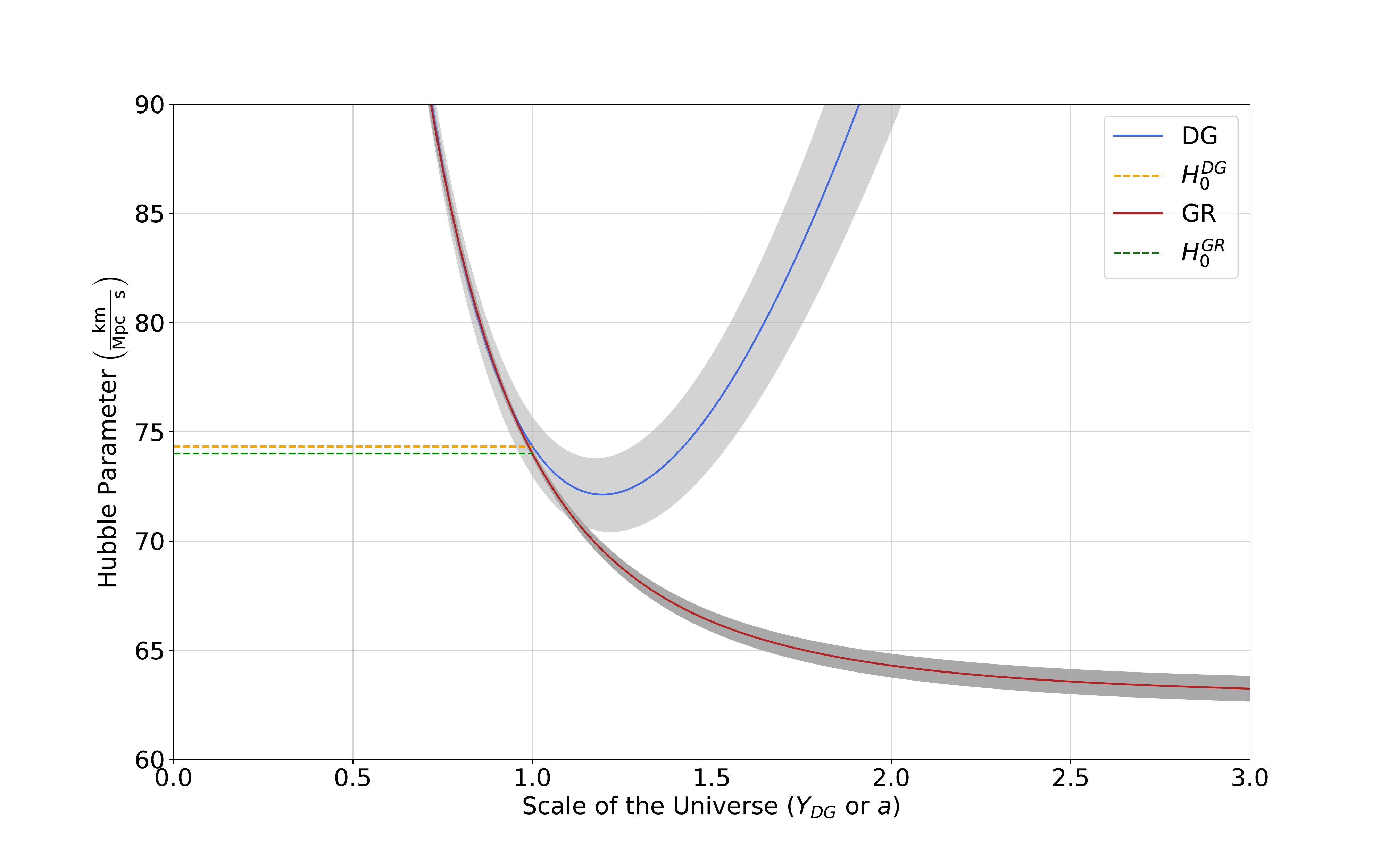}
\caption{Hubble Parameter for DG and GR fitted models assuming $M = -19.23$}
\label{Fig:HvsScaleFactor}
\end{figure}

The Table \ref{tab:H0compared} shows that the DG prediction for $H_0$ is in concordance with the last $H_0$ measurement, we are interested in preserve this measurement because we want to make SNe-Ia and CMB compatible. \footnote{``The direct measurement is very model-independent, but prone to systematics related to local flows and the standard candle assumption. On the other hand, the indirect method is very robust and precise, but relies completely on the underlying model to be correct. Any disagreement between the two types of measurements could in principle point to a problem with the underlying $\Lambda$CDM model.'' (\cite{Odderskov})} This compatibility is a consequence of the excellent fit (we are only working with $h$ and $L_2$) and the series expansion of the $d_L^{DG}$ in terms of $z$ (This term can be expanded as a $z$ series, with the same physical significance, such as the Hubble Constant and the deceleration parameter, but these parameters depend in a very different form compared to GR). GR also predicts a high $H_0$ value with the same assumptions, but it needs to include $\Lambda$ to fit the SN-Ia data. The last two data labeled as GR and DG in the Table are related to the full SNe-Ia data set, and not with a local measurement. \medskip

The Figure \ref{Fig:HvsScaleFactor} shows the change in the Hubble parameter for both models. In the DG case, the Hubble parameter increases after $Y_{DG} \approx 1.2$, and the Universe starts to increases its size to end with a Big Rip. In contrast, as we know, LCDM does not predict a Big Rip. The $H(a)$ tends to be constant when $a\to \infty$ (\cite{universe5020051}).

\begin{figure}[h!]
\centering
\includegraphics[width=1\textwidth]{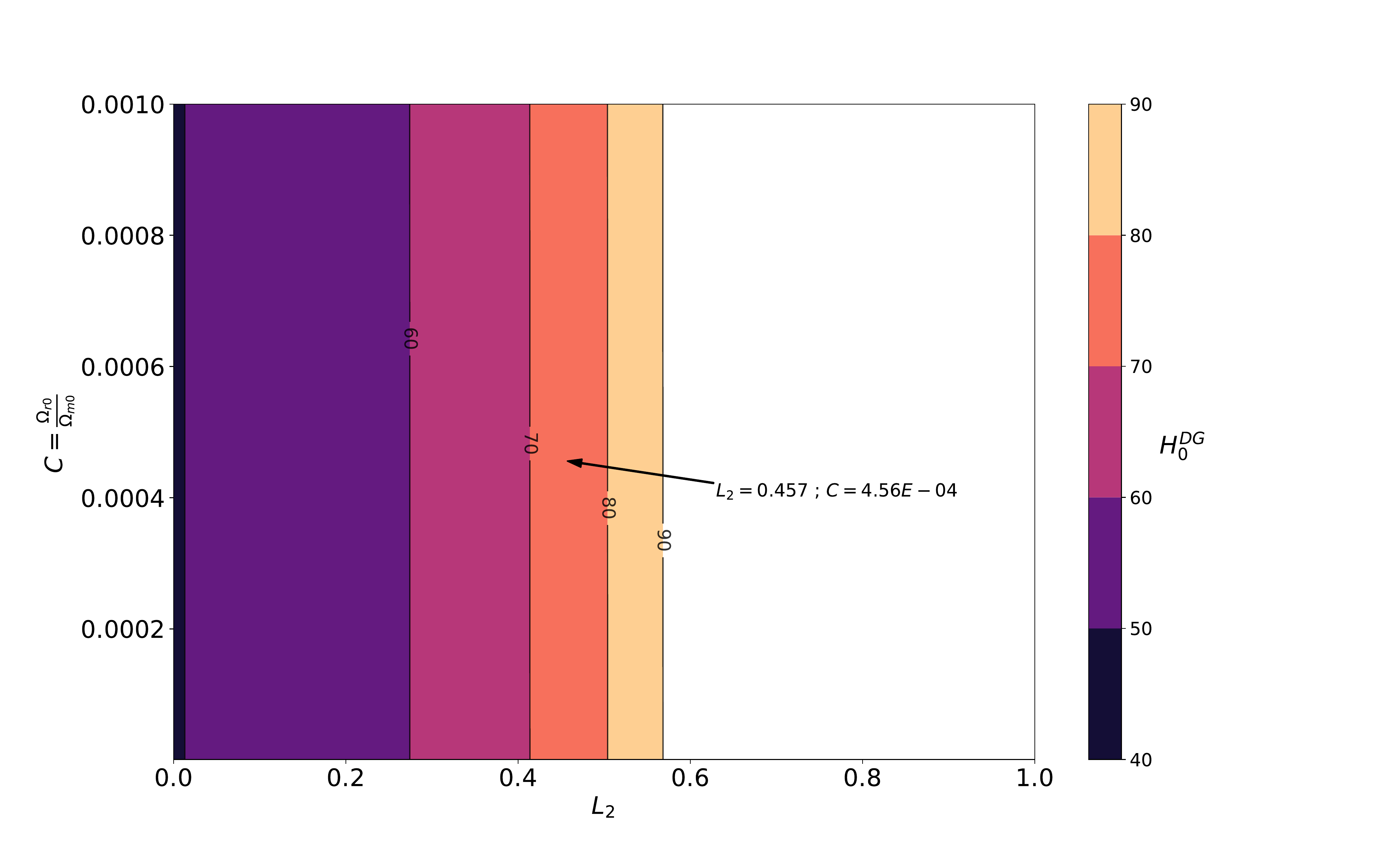}
 \caption{Dependence of the Hubble Parameter for DG with $C$ and $L_2$.}
 \label{Fig:H0Contourzoom}
\end{figure}

The Figure \ref{Fig:H0Contourzoom} shows how the deceleration parameter depends on $C$ and $L_2$. In the regime of interest, where $C\to 10^{-4}$, $H_0^{DG}$, $q_0$ is independent of $C$ and it increases with $L_2$.

In GR, the deceleration parameter is calculated from the Equation \eqref{GRqparameter} and the Friedmann equations

\begin{equation}
{\displaystyle q_{0}={\frac {1}{2}}\Omega _{m,0}-\Omega _{\Lambda,0 }.}
\label{q0_GR_Omega_expresion}
\end{equation}

For DG, we used the Equation \eqref{DGqparameter}. To evaluate the deceleration today, we evaluate $a=1$ for GR, and $Y=1$ for DG.

We show the Deceleration Parameters for both models in the Table \ref{tableq0compared}. Both models have $q_0<0$; in other words, the Universe is accelerating but with slightly different rates.

\begin{figure}
\centering
\includegraphics[width=16cm,keepaspectratio]{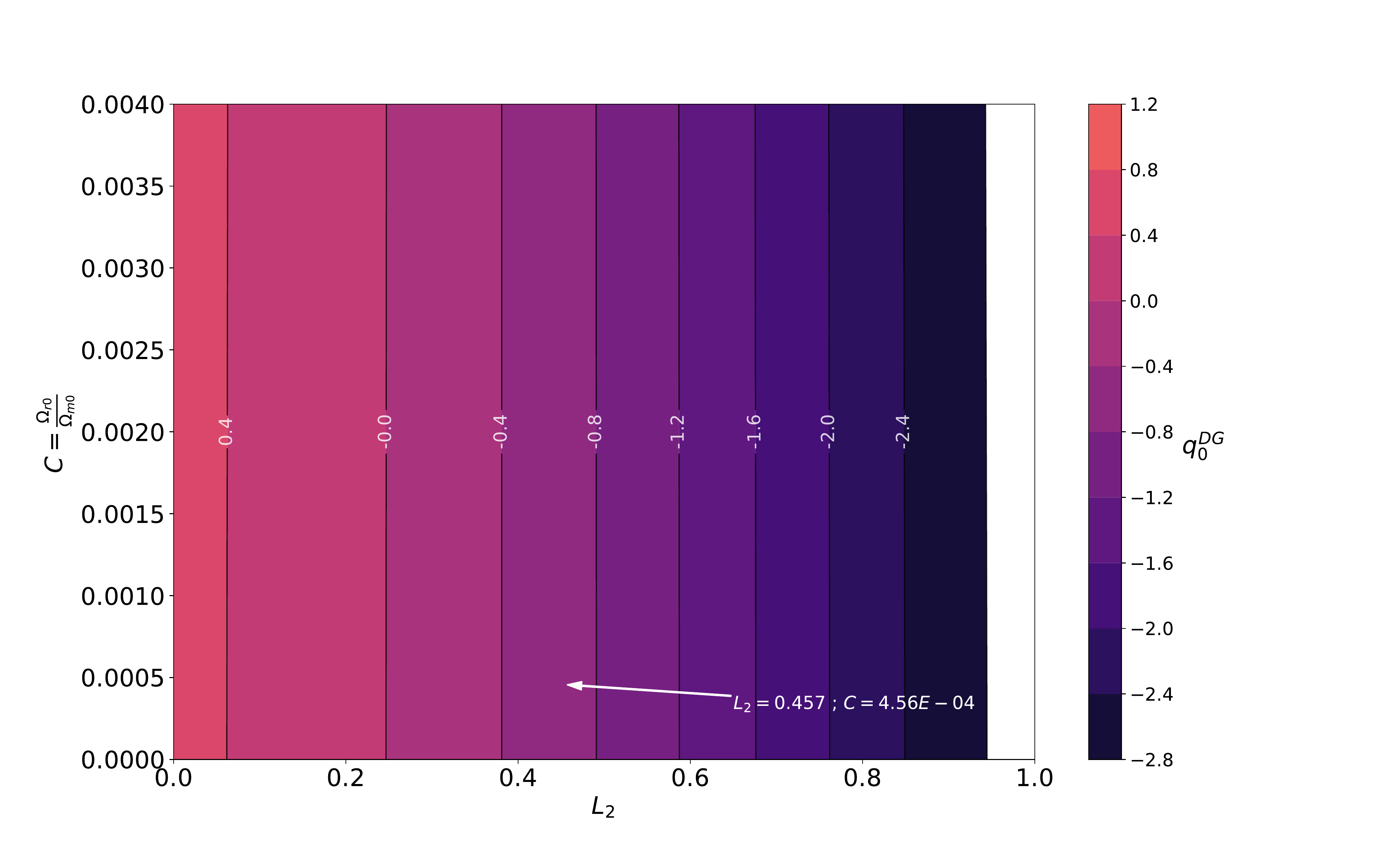}
\caption{The Figure shows the dependence of the deceleration Parameter for DG with $C$ and $L_2$.}
\label{Fig:q0_contoursplot_zoom}
\end{figure}

In the Figure \ref{Fig:q0_contoursplot_zoom} we show how the deceleration parameter depends on $C$ and $L_2$. It is important consider that acceleration depends on $L_2$  and it is independent of $C$ (if $C$ is small). 

\begin{table}
\centering
\begin{tabular}{l*{6}{c}r}
Model     & $q_0$  & Error \\
\hline
DG & -0.700 & 0.001  \\
GR & -0.58 & 0.02 
\end{tabular}
\caption{$q_0$ values were found using Least Squares Method with SNe-Ia data.}
\label{tableq0compared}
\end{table}

The $L_2$ parameter is driving the acceleration, and it is describing the SNe-Ia data. If $L_2 \to 1$, then $q_0$ is more negative, and the Universe has a higher acceleration.\medskip

\subsubsection{Cosmic Time and redshift}

To calculate the Cosmic Time in DG, we used the Equation \eqref{CosmologicalTimeDeltaGravity}. The redshift is obtained by numerical solution from the Equation \eqref{Ytilderedshift}. 

Meanwhile, for the GR model, we obtained the Cosmic Time integrating the first Friedmann equation and solving $t(\Omega_{m,0},H_0)$. Here we have included  $\Omega_{\Lambda}=1-\Omega_{m,0}$ and we chose a flat cosmology and $\Omega_{r,0}=0$. The integral for the first Friedmann equation can be analytically solved (from the Equation  \ref{FirstFriedmannEquation2}):

\begin{equation}
t=\int_0^a\frac{1}{\sqrt[]{\frac{\Omega_{m,0}}{x}+(1-\Omega_{m,0})x^2}}dx = \frac{2}{3\sqrt[]{1-\Omega_{m,0}}} \ln\left(\frac{\sqrt[]{-\Omega_{m,0}a^3+\Omega_{m,0}+a^3}+\sqrt[]{1-\Omega_{m,0}}a^{3/2}}{\sqrt[]{\Omega_{m,0}}}\right),
\label{CosmicTimeGR0}
\end{equation}

where $t$ in \eqref{CosmicTimeGR0} is the Cosmic Time for GR. The behavior of Cosmic Time dependence with redshift for both models is very similar (\cite{universe5020051}).

The age of the Universe in DG is calculated using the Equation \eqref{CosmologicalTimeDeltaGravity}. $t(Y)$ only depends on $h$ and $C$, but not on $L_2$. To calculate the age of the Universe in DG, we evaluate $Y=1 \gg C$, then the age only depends on $h$. On the other hand, in GR we calculate the age of the Universe we use the Equation \eqref{CosmicTimeGR0} that requires $h$ and $\Omega_{m,0}$. The age for DG model is  $13.1 \pm 0.1$ Gyrs and for GR is  $13.0 \pm 0.2$ Gyrs. \medskip
 
The higher the Hubble Constant, the lower the age of the Universe. This relation is vital since if the local fit of supernovae radically changes $H_0$, then the age of the Universe changes.

The age of the Universe for DG and GR are small (13.1 Gyrs for DG and  13.0 Gyrs for GR) compared with the age calculated from Planck (13.8 Gyrs). A crucial and precise estimation made by \cite{refId0} based on the ages of globular clusters in the Milky Way (which is independent of cosmology) indicates that the Universe has to be older than 13.6 $\pm$ 0.8 Gyrs. DG, assuming the results of SNe-Ia local measurements, is on the verge of this observational constraint. We emphasize that the problem goes beyond DG because this discrepancy is related to the local measurements and it is due to the calibration made by \cite{Riess2016}. As we discussed in the Section \ref{sec:intro}, there are many different $H_0$ measurements, but in this work we are working assuming that this high $H_0$ value is correct. \medskip

\section{TT CMB scalar spectrum}\label{sec:CMB}

To fit the CMB power spectrum with DG we have to use perturbation theory. The perturbation theory has been developed in previous work (\cite{Alfaro:2020qbf}), where we have decomposed the perturbation terms as the standard Scalar-Vector-Tensor method. Here we show a summary of the main equations required to obtain the CMB and fit the parameters. The metric is perturbed up to first order\footnote{For a full development about the DG perturbation theory, the reader can visit the preprint in \url{https://arxiv.org/abs/2001.08354}.}:

\begin{align}
g_{\mu\nu}=\bar g_{\mu\nu}+h_{\mu\nu},\\
\tilde{g}_{\mu\nu}=\tilde{\bar{g}}_{\mu\nu}+\tilde{h}_{\mu\nu}.
\end{align}

In particular, we followed the Weinberg's approach \cite{weinberg2008cosmology} (he developed this method in the synchronous gauge \footnote{There are other methods, to solve the equations in an analytic approach, assuming some approximations, \cite{Mukhanov2004,weinberg2008cosmology}.}), which consist in two main aspects: the first one is the so-called hydrodynamic limit, which consists on that near recombination time photons were in local thermal equilibrium with the baryonic plasma, then photons can be treated hydro-dynamically, like plasma and cold dark matter. The second assumption is a sharp transition from thermal equilibrium to complete transparency at last scattering moment $t_L$. \medskip

In this context, the components of the Universe are photons, neutrinos, baryons, and cold dark matter and the Delta sector. The approximation used here neglected both anisotropic energy-momentum tensors and assumed the usual equation of state for the components. Besides, as we treated photons and  Delta photons hydro-dynamically, we used $\delta u_{\gamma}=\delta u_B$ and $\delta \tilde{u}_{\gamma}=\delta \tilde{u}_B$ (velocity perturbations). Moreover, as the synchronous scheme did not fully fix the gauge, the remaining degree of freedom were used to fix $\delta u_D=0$, which means that cold dark matter evolves at rest with respect to the Universe expansion. In our theory, the extended synchronous scheme also had an extra degree of freedom, which we used to put $\delta\tilde{u}_D=0$ as its standard part.
 
It is useful to rewrite these equations in terms of the following dimensionless term:
  
 \begin{equation}\label{dim frac pert}
\delta_{\alpha q}=\frac{\delta \rho_{\alpha q}}{\bar{\rho}_{\alpha}+\bar{p}_{\alpha}}\,,
   \end{equation}
 
 where $\alpha$ can be $\gamma$, $\nu$, $B$ and $D$ (photons, neutrinos, baryons and dark matter, respectively) and $q$ is the mode. Also we used $R=3\bar{\rho}_B/4\bar{\rho}_{\gamma}$ and $\tilde{R}=3\tilde{\bar{\rho}}_D/4\tilde{\bar{\rho}}_{\gamma}$. By the other side, in the Delta sector we used a dimensionless fractional perturbation. However, this perturbation was defined as the Delta transformation of Equation \eqref{dim frac pert} \footnote{We choose this definition because the system of equations now seems as an homogeneous system exactly equal to the GR sector (where now the variables were the Delta-fields) with external forces mediated by the GR solutions. Maybe the most intuitive solution should be $$\tilde{\delta}^{int}_{\alpha q}=\frac{\delta\tilde{\rho}_{\alpha q}}{\tilde{\bar{\rho}}_{\alpha}+\tilde{\bar{p}}_{\alpha}}\;,$$ however these definitions are related by $$\tilde{\delta}_{\alpha q}=\frac{\tilde{\bar{\rho}}_{\alpha}+\tilde{\bar{p}}_{\alpha}}{\bar{\rho}_{\alpha}+\bar{p}_{\alpha}}\left(\tilde{\delta}^{int}_{\alpha q}-\delta_{\alpha q}\right)\;.$$},
 
 \begin{equation}\label{tilde dim frac pert}
\tilde{\delta}_{\alpha q}\equiv \tilde{\delta} \delta_{\alpha q}=\frac{\delta \tilde{\rho}_{\alpha q}}{\bar{\rho}_{\alpha}+\bar{p}_{\alpha}}-\frac{\tilde{\bar{\rho}}_{\alpha}+\tilde{\bar{p}}_{\alpha}}{\bar{\rho}_{\alpha}+\bar{p}_{\alpha}}\delta_{\alpha q}\,.
 \end{equation}
 
 The equations for the GR sector are
 
 \begin{eqnarray}\label{GRequations}
\frac{d}{dt}\left(a^{2}\dot{\Psi}_{q}\right) & = & -4\pi Ga^{2}\left(\bar{\rho}_{D}\delta_{Dq}+\bar{\rho}_{B}\delta_{Bq}+\frac{8}{3}\bar{\rho}_{\gamma}\delta_{\gamma q}+\frac{8}{3}\bar{\rho}_{\nu}\delta{}_{\nu q}\right)\;, \label{final eq1}\\
\dot{\delta}_{\gamma q}-(q^2/a^{2})\delta u_{\gamma q} & = & -\dot{\Psi}_{q}\; ,\label{final fot}\\
\dot{\delta}_{Dq} & = & -\Psi_{q}\;,\\
\dot{\delta}_{Bq}-(q^2/a^{2})\delta u_{\gamma q} & = & -\dot{\Psi}_{q}\; ,\label{final bar}\\
\dot{\delta}_{\nu q}-(q^2/a^{2})\delta u_{\nu q} & = & -\dot{\Psi}_{q}\; ,\\
\frac{d}{dt}\left(\frac{\left(1+R\right)\delta u_{\gamma q}}{a}\right) & = & -\frac{1}{3a}\delta_{\gamma q}\; ,\\
\frac{d}{dt}\left(\frac{\delta u_{\nu q}}{a}\right) & = & -\frac{1}{3a}\delta_{\nu q}.\label{final eq2}
\end{eqnarray}
 
 While, the equations for the DG sector are

 \begin{eqnarray} \label{fdelta eq1}
\left[2\dot{F}\frac{\dot{a}}{a}+\ddot{F}\right]a^2\Psi_q+\left[6F\frac{\dot{a}}{a}+\frac{5}{2}\dot{F}\right]a^2\dot{\Psi}_q+3Fa^2\ddot{\Psi}_q-\frac{d}{dt}\left(a^2\dot{\tilde{\Psi}}_q\right)&=&\frac{\kappa}{2} a^2\left[\bar{\rho}_D\tilde{\delta}_{Dq}\right.\nonumber\\+\left.\bar{\rho}_B\tilde{\delta}_{Bq}+\frac{8}{3}\bar{\rho}_{\gamma}\tilde{\delta}_{\gamma q}+\frac{8}{3}\bar{\rho}_{\nu}\tilde{\delta}_{\nu q}-\frac{F}{2}\left(\bar{\rho}_D{\delta}_{Dq}+\bar{\rho}_B{\delta}_{Bq}\right)-\frac{8}{3}F\left(\bar{\rho}_{\gamma}{\delta}_{\gamma q}+\bar{\rho}_{\nu}{\delta}_{\nu q}\right)\right],
  \end{eqnarray}

\begin{eqnarray}
\dot{\tilde{\delta}}_{\gamma q}-\frac{q^2}{a^2}\left(\delta\tilde{u}_{\gamma q}+F\delta u_{\gamma q}\right)+\dot{\tilde{\Psi}}_q-\partial_0(F\Psi_q)&=&0\;,\label{final delta fot}\\
\dot{\tilde{\delta}}_{Dq}+\dot{\tilde{\Psi}}_q-\partial_0(F\Psi_q)&=&0\;,\\
\dot{\tilde{\delta}}_{Bq}-\frac{q^2}{a^2}\left(\delta\tilde{u}_{\gamma q}+F\delta u_{\gamma q}\right)+\dot{\tilde{\Psi}}_q-\partial_0(F\Psi_q)&=&0\;,\label{final delta B}\\
\dot{\tilde{\delta}}_{\nu q}-\frac{q^2}{a^2}\left(\delta\tilde{u}_{\nu q}+F\delta u_{\nu q}\right)+\dot{\tilde{\Psi}}_q-\partial_0(F\Psi_q)&=&0\;,\\
\frac{\tilde{\delta}_{\gamma q}}{3a}+\frac{d}{dt}\left(\frac{(1+R)\delta \tilde{u}_{\gamma q}}{a}\right)+2F\frac{d}{dt}\left(\frac{(R-\tilde{R})\delta {u}_{\gamma q}}{a}\right)-F\frac{d}{dt}\left(\frac{(1+R)\delta u_{\gamma q}}{a}\right)\nonumber\\-2\dot{F}(\tilde{R}-R)\frac{\delta u_{\gamma q}}{a}&=&0\;,\\
\frac{\tilde{\delta}_{\nu q}}{3a}+\frac{d}{dt}\left(\frac{\delta\tilde{u}_{\nu q}}{a}\right)-F\frac{d}{dt}\left(\frac{\delta u_{\nu q}}{a}\right)&=&0\;, \label{fdelta eq2}
\end{eqnarray}

where $\Psi_q$ and $\tilde{\Psi}_q$ are a particular combination of the scalar perturbations in the metric (\cite{Alfaro:2020qbf}).

\subsection{Matter era}\label{matter era}

In this era $a\gg C$ \footnote{ and $R=\tilde{R}=0$.}, and the perturbative equations for GR can be approximated and solved. These solutions are given by \footnote{${\cal R}_q$ is defined as $
q^{2}{\cal R}_{q}\equiv-a^{2}H\Psi_{q}+4\pi Ga^{2}\delta\rho_{q}+q^{2}H\delta u_{q} $. It is a gauge invariant quantity, which take a time independent value for $q/a\ll H$. \cite{weinberg2008cosmology} }

\begin{eqnarray}\label{GR pert solutions 1}
  \delta_{Dq}&=&\frac{9q^2t^2{\cal R}_q{\cal T}(\kappa)}{10a^2},\\
  \dot{\Psi}_q&=&-\frac{3q^2t{\cal R}_q{\cal T}(\kappa)}{5a^2}\label{GR pert solutions 2},
\end{eqnarray}

\begin{eqnarray}\label{GR pert solutions 3}
  \delta_{\gamma q}=\delta_{\nu q}=\frac{3{\cal R}_q}{5}\left[{\cal T}(\kappa)-{\cal S}(\kappa)\cos\left(q\int_0^t\frac{dt}{\sqrt{3}a}+\Delta(\kappa)\right)\right]\;,\\
  \delta u_{\gamma q}=\delta u_{\nu q}=\frac{3t{\cal R}_q}{5}\left[-{\cal T}(\kappa)+{\cal S}(\kappa)\frac{a}{\sqrt{3}qt}\sin\left(q\int_0^t\frac{dt}{\sqrt{3}a}+\Delta(\kappa)\right)\right]\label{GR pert solutions 3.5},
\end{eqnarray}

where ${\cal T}(\kappa)$, ${\cal S}(\kappa)$ and $\Delta(\kappa)$ are known as transfer functions. They only depend on 

\begin{equation}\label{kappa}
\kappa \equiv \frac{q\sqrt{2}}{a_{EQ}H_{EQ}},
\end{equation}

where $a_{EQ}$ and $H_{EQ}$ are, the Scale Factor and the expansion rate at the matter-radiation equality (\cite{weinberg2008cosmology}).\medskip 

To get all the transfer functions, we have to compare solutions with the full equation system (with $\rho_B=\tilde{\rho}_B=0$). To do this, we define $y\equiv a/a_{EQ}=a/C$ and use the following changes of variables:

\begin{equation}\label{change of variable}
\frac{d}{dt}=\frac{H_{EQ}}{\sqrt{2}}\frac{\sqrt{1+y}}{y}\frac{d}{dy},   
\end{equation}

\begin{eqnarray*}
  \delta_{Dq}=\kappa^2{\cal R}^0_q d(y)/4\,\,,\;\;\;\;\delta_{\gamma q}=\delta_{\nu q}=\kappa^2{\cal R}^0_qr(y)/4\;,\\
  \dot{\Psi}_q=(\kappa^2H_{EQ}/4\sqrt{2}){\cal R}^0_qf(y)\,\,,\;\;\;\delta u_{\gamma q}=\delta u_{\nu q}=(\kappa^2\sqrt{2}/4H_{EQ}){\cal R}^0_q g(y)\;,
\end{eqnarray*}

\begin{eqnarray*}
  \tilde{\delta}_{Dq}=\kappa^2{\cal R}^0_q\tilde{d}(y)/4\,\,,\;\;\;\;\tilde{\delta}_{\gamma q}=\tilde{\delta}_{\nu q}=\kappa^2{\cal R}^0_q\tilde{r}(y)/4,\\
  \dot{\tilde{\Psi}}_q=(\kappa^2H_{EQ}/4\sqrt{2}){\cal R}^0_q\tilde{f}(y)\,\,,\;\;\;\delta\tilde{u}_{\gamma q}=\delta\tilde{u}_{\nu q}=(\kappa^2\sqrt{2}/4H_{EQ}){\cal R}^0_q\tilde{g}(y)\;.
\end{eqnarray*}

Then, the perturbative equations given in the matter era for GR and DG can be rewritten as

\begin{eqnarray}
  \sqrt{1+y}\frac{d}{dy}\left(y^2{f}(y)\right)&=&-\frac{3}{2}{d}(y)-\frac{4{r}(y)}{y},\label{Gr eq 1}\\
  \sqrt{1+y}\frac{d}{dy}{r}(y)-\frac{\kappa^2{g}(y)}{y}&=&-y{f}(y),\\
  \sqrt{1+y}\frac{d}{dy}{d}(y)&=&-y{f}(y),\\
  \sqrt{1+y}\frac{d}{dy}\left(\frac{{g}(y)}{y}\right)&=&-\frac{{r}(y)}{3}\,\label{Gr eq 2},
\end{eqnarray}

and

\begin{eqnarray}
-\left[(1+2y)yF'(y)+y(1+y)F''(y)\right]d(y)+\left[6F(y)+\frac{5}{2}yF'(y)\right]y\sqrt{1+y}f(y)\nonumber\\+3F(y)y^2\sqrt{1+y}f'(y)-\sqrt{1+y}\frac{d}{dy}\left(y^2\tilde{f}(y)\right)=\frac{3\tilde{d}(y)}{2}+\frac{4\tilde{r}(y)}{y}\nonumber\\-\frac{3F(y)d(y)}{4}-\frac{4F(y)r(y)}{y}\label{Dg eq 1},\\
\sqrt{1+y}\frac{d}{dy}\tilde{d}(y)=-y\tilde{f}(y)-\sqrt{1+y}\frac{d}{dy}d(y),\\
\sqrt{1+y}\frac{d}{dy}\tilde{r}(y)=\frac{\kappa^2}{y}[\tilde{g}(y)+F(y)g(y)]-y\tilde{f}(y)-\sqrt{1+y}\frac{d}{dy}d(y),\\
\sqrt{1+y}\frac{d}{dy}\left(\frac{\tilde{g}(y)}{y}\right)=-\frac{\tilde{r}(y)}{3}+\sqrt{1+y}F(y)\frac{d}{dy}\left(\frac{g(y)}{y}\right)\label{Dg eq 2}\,.
\end{eqnarray}

Now, we have to calculate the initial condition-behavior described by the radiation-dominated era (we have to approximate the original equations in this regime). In other words, at the beginning of the matter-dominated era, we have the following initial conditions

\begin{eqnarray*}
  d(y) = r(y)\rightarrow y^2, \\
  f(y)\rightarrow -2, \\
  g(y)\rightarrow -\frac{y^4}{9},
\end{eqnarray*}

\begin{eqnarray*}
\tilde{d}(y)=\tilde{r}(y)\rightarrow -\frac{L_2C^{3/2}}{3}y^3,\\
\tilde{f}(y)\rightarrow\sqrt{2}L_2C^{3/2}y,\\
\tilde{g}(y)\rightarrow\frac{L_2C^{3/2}}{2}y^5.
\end{eqnarray*}

Now, we have to include the $ R $ and $\tilde{R}$ factors that were not considered as a part of the equations.  This step was done with WKB approximation (\cite{weinberg2008cosmology}). Also, we have to include the damping effect in the fluid of baryons and photons. This effect is known as {\it Silk damping} and considers coefficients of shear viscosity, heat conduction, bulk viscosity, and Thomson scattering associated with the fluid (\cite{1983MNRAS.202.1169K,Silk:1967kq,weinberg_fluid}). Then the full solutions for the photon density perturbations are

\begin{eqnarray}\label{GR pert solutions 4}
  \delta_{\gamma q}&=&\frac{3{\cal R}_q^o}{5}\left[{\cal T}(\kappa)(1+3R)\right.\nonumber\\&&\left.-(1+R)^{-1/4}e^{-\int_0^t\Gamma dt}{\cal S}(\kappa)\cos\left(\int_0^{t}\frac{qdt}{\sqrt{3(1+R(t))}a_{DG}(t)}+\Delta(\kappa)\right)\right]\,,\\
  \delta u_{\gamma q}&=&\frac{3{\cal R}_q^o}{5}\left[-t{\cal T}(\kappa)\right.\nonumber\\&&+\left.\frac{a_{DG}}{\sqrt{3}q(1+R)^{3/4}}e^{-\int_0^t\Gamma dt}{\cal S}(\kappa)\sin\left(\int_0^{t}\frac{qdt}{\sqrt{3(1+R(t))}a_{DG}(t)}+\Delta(\kappa)\right)\right]\,,
\end{eqnarray}

where 

\begin{equation}
    \Gamma= \frac{q^2t_{\gamma}}{6a_{DG}^2(1+R)}\left[\frac{16}{15}+\frac{R^2}{1+R}\right],
\end{equation}

where $t_{\gamma}$ is the mean free time for photons. We remark that at this level, we used $a\sim a_{DG}$ because these solutions are valid when DG approaches to GR at the beginning of the Universe. In particular, those solutions at the moment of the last scattering play a crucial role when we compute the temperature multipole coefficients.\medskip

Now we have to express the temperature's perturbation as a function of the densities perturbations. This procedure is long and takes many pages. It is not the objective of this paper to show the steps to obtain this result (\cite{Alfaro:2020qbf}). However, it is vital to understand the physics behind the equations, the approximations, and the numerical contributions behind every term. First of all, we show four essential functions called Form Factors that are the main contributions to the TT CMB spectrum,

\begin{eqnarray}
  {\cal F}(q)&=&-\frac{1}{2}a_{DG}^2(t)\ddot{B}_q(t_{ls})-\frac{1}{2}a_{DG}(t)\dot{a}_{DG}(t_{ls})\dot{B}_q(t_{ls})+\frac{1}{2}E_q(t_{ls})+\frac{\delta T_q(t_{ls})}{\bar{T}(t_{ls})},\label{formfactorfeo0}\\
  \tilde{{\cal F}}(q)&=&-\frac{1}{2}a_{DG}^2(t)\ddot{\tilde{B}}_q(t_{ls})-\frac{1}{2}a_{DG}(t_{ls})\dot{a}_{DG}(t_{ls})\dot{\tilde{B}}_q(t_{ls}),\\
  {\cal G}(q)&=&-q\left(\frac{1}{2}a_{DG}(t_{ls})\dot{B}_q(t_{ls})+\frac{1}{(1+3F(t_{ls}))a_{DG}(t_{ls})}\delta u_{\gamma}(t_{ls})\right),\\
  \tilde{{\cal G}}(q)&=&-q\left(\frac{1}{2}a_{DG}(t_{ls})\dot{\tilde{B}}_q(t_{ls})+\frac{1}{(1+3F(t_{ls}))a_{DG}(t_{ls})}\delta \tilde{u}_{\gamma}(t_{ls})\right)\label{formfactorfeo1}\,.
\end{eqnarray}

where the TT CMB spectrum is given by the Equation \eqref{final multipole formula}. These formulas will be very useful. \footnote{The $B_q$, $\tilde{B}_q$ and $E_q$ are scalar perturbative terms that appears in the SVT decomposition. For more details please see the preprint in \url{https://arxiv.org/abs/2001.08354}}

These Form Factors can be rearranged using many new definitions that introduce physics notation. Before doing that, it is important to define some physical concepts.\medskip

\paragraph{Angular distance \texorpdfstring{$d_A^{DG}$}{dA DG}}

The relation between the luminosity distance and angular distance expressed by the Equation \eqref{dLdARelation} in DG is preserved and we used it to find the $d_A^{DG}$ at a given redshift. In the DG perturbative equations, the angular distance appears naturally as $d_A(t_{ls}) = r_{ls}  a_{DG}(t_{ls})$. This equation is the same definition given here, evaluated at the Last Scattering surface.  The angular distance is crucial to define the physical meaning of the next equations.\medskip

In equations,

\begin{eqnarray}
  d_A^{DG}(t_{ls})&=&ca_{DG}(t_{ls})\int_{t_{ls}}^{t_0}\frac{dt'}{a_{DG}(t')}=c\frac{a_{DG}(t_0)}{1+z_{ls}}\int_{t_{ls}}^{t_0}\frac{dt'}{a_{DG}(t')}=c\frac{1}{1+z_{ls}}\int_{t_{ls}}^{t_0}\frac{dt'}{Y_{DG}(t')}\\&=&c\frac{1}{1+z_{ls}}\int_{Y_{ls}}^{1}\frac{dY'}{Y_{DG}(Y')}\frac{dt}{dY'}=\frac{d_L^{DG}(t_{ls})}{(1+z_{ls})^2}.
\end{eqnarray}

\paragraph{Horizon distance \texorpdfstring{$d_H^{DG}$}{dH DG}}

We have to consider the effective metric. This will produce the same integrand as the Equation \eqref{LuminosityDistanceDG} but substituting $a(t) \to Y_{DG}(Y)$. Note that $Y_{DG}$ depends on $Y(t)$. We have to apply the chain rule and also change the integral limits to  $\int_0^{Y(z)}$. Finally, the Horizon distance in DG is given by

\begin{equation}
d_H^{DG}(z,L_2,C)=\frac{\sqrt{1+C}}{(1+z) 100h} \int_{0}^{Y(z)}c_s \frac{Y}{\sqrt{Y+C}}\frac{dY}{{Y_{DG}}}.
\label{HorizonDistanceDG}
\end{equation}

Note 1: The speed of light $c$ has been replaced by $c_s$, where the subscript $s$ represents the sound. This change is introduced because we want to use this equation to calculate the acoustic horizon distance. This acoustic horizon is the maximum distance that a fluid with speed $c_s$ has traveled between redshift $\in (\infty, z)$.

Note 2: Do not confuse $C$ in terms of GR densities that are not physical with physical densities labeled with $^{DG}$ or $_{DG}$. For example, $h^2\Omega_{r,0}$ is not a physical density.\medskip

In the standard cosmology, the speed of sound is given by

\begin{equation}
    c_s^2 = \frac{\delta p}{\delta \rho} = \frac{1}{\sqrt{3(1+R)}},
    \label{c_s}
\end{equation}

where $R = \frac{4\rho_b}{3\rho_\gamma}$ in GR. We emphasize that Delta matter and Delta radiation could change this equation. In the simplest case, Delta particles do not affect the speed of sound of the fluid because we are assuming that Delta particles behave like dark matter particles: they are non-interacting particles. Neither dark matter appears in this equation nor the Delta particles. However, in DG we use the following definition:

\begin{equation}
R = \frac{4h^2\Omega_{b}^{DG}}{3h^2\Omega_{\gamma}^{DG}}.
\label{speed_equation_DG}
\end{equation}

Now, $R$ is a function of physical densities. We did not include the Delta matter or the Delta radiation.

Unfortunately, due to all the approximations we have used, we need to add one more correction to the GR sector's solutions. We considered a sharp transition from the moment when the Universe was opaque to transparent. However, this was not instantaneous, yet it could be considered gaussian. This normal distribution implies an effect known as Landau damping (\cite{Landau:437300}), and it is related to the dispersion of the distribution of a wavefront in a plasma. This consideration is relevant, and it is related to the standard deviation of temperature at the Last Scattering moment (labeled as $ls$). With these considerations, the solutions of the perturbations are given by:

\begin{eqnarray}
  \dot{\Psi}_q(t_{ls})&=&-\frac{3q^2t_{ls}{\cal R}_q^o{\cal T}(\kappa)}{5a_{DG}^2(t_{ls})}\,,\\
  \delta_{\gamma q}(t_{ls})&=&\frac{3{\cal R}_q^o}{5}\left[{\cal T}(\kappa)(1+3R_{ls})-(1+R_{ls})^{-1/4}e^{-q^2d^2_D/a^2_{DG}(t_{ls})}\right.\nonumber\\&\times&\left.{\cal S}(\kappa)\cos\left(q\int_0^{t_{ls}}\frac{dt}{\sqrt{3(1+R(t))}a_{DG}(t)}+\Delta(\kappa)\right)\right]\,,\\
  \delta u_{\gamma q}(t_{ls})&=&\frac{3{\cal R}_q^o}{5}\left[-t_{ls}{\cal T}(\kappa)+\frac{a_{DG}(t_{ls})}{\sqrt{3}q(1+R_{ls})^{3/4}}e^{-q^2d^2_D/a_{DG}^2(t_{ls})}\right.\nonumber\\&\times&\left.{\cal S}(\kappa)\sin\left(q\int_0^{t_{ls}}\frac{dt}{\sqrt{3(1+R(t))}a_{DG}(t)}+\Delta(\kappa)\right)\right]\,,
\end{eqnarray}

where

\begin{eqnarray}\label{dampings}
  d^2_D&=&d^2_{Silk}+d^2_{Landau} \label{full_damp}\,,\\
  d^2_{Silk}&=&Y_{DG}^2(t_{ls})\int_0^{t_{ls}}\frac{t_{\gamma}}{6Y_{DG}^2(1+R)}\left\{\frac{16}{15}+\frac{R^2}{(1+R)}\right\}dt\,,\label{Silk_damp}\\
  d^2_{Landau}&=&\frac{\sigma_t^2}{6(1+R_{ls})} \label{Landau_damp}\,,
\end{eqnarray}

and $t_{\gamma}$ is the mean free time for photons and $R=3\bar{\rho}_B^{DG}/4\bar{\rho}_{\gamma}^{DG}=3h^2\Omega_{b,0}^{DG}Y_{DG}/4h^2\Omega_{\gamma,0}^{DG}$. The $d$ notation characterizes the damping length for each damping process.\\
In order to evaluate the Silk damping, we use

\begin{equation}
t_{\gamma}=\frac{1}{n_e\sigma_Tc},
\end{equation}

where $n_e$ is the number density of electrons, and $\sigma_T$ is the Thomson cross-section. On the other hand

\begin{eqnarray}
    q\int_0^{r_{ls}} c_s dr&=& q\int_0^{t_{ls}}\frac{dt}{\sqrt{3(1+R(t))}a_{DG}(t)}\equiv qr^{SH}_{ls}\nonumber\\&=&\frac{q}{a_{DG}(t_{ls})}\cdot (a_{DG}(t_{ls})r_{ls}^{SH})=\frac{q}{a_{DG}(t_{ls})}\cdot d_H(t_{ls})
\end{eqnarray}

where $c_s$ is the speed of sound, $r^{SH}_{ls}$ is the sound horizon radial coordinate and $d_H$ is the horizon distance, and $\kappa=qd_T^{DG}/a_{DG}(t_{ls})$ (defined in Equation \eqref{kappa}) implies

\begin{equation}
d_T^{DG}(t_{ls})\equiv c\frac{\sqrt{2} a_{DG}(t_{ls})}{a_{EQ}H_{EQ}}  =c \frac{a_{DG}(t_{ls})\sqrt{\Omega_R}}{H_{0}\Omega_M}= c \frac{a_{DG}(t_{ls})}{100h}\sqrt{C(C+1)}.
\label{kappa_dT_transformation}
\end{equation}

We must include that, in $z_{reion} \sim 10$ (reionization), the neutral hydrogen left over from the time of recombination becomes reionized by ultraviolet light from the first generation of massive stars (\cite{weinberg2008cosmology,PiattellaNotes}).
The photons of the cosmic microwave background have a small but non-negligible probability $1 - exp(-\tau_{reion})$ (where $\tau_{reion}$ is the optical depth of the reionized plasma) of being scattered by the electrons set free by this reionization. The TT spectrum is a quadratic function of the  the temperature fluctuations, then we have to weigh the spectrum by a factor $exp(-2\tau_{reion})$\footnote{ In the standard GR case, the observations from polarization spectrum suggests that $exp(-2\tau_{reion})\approx 0.8$. We used this value to fit the spectrum. We did not study the reionization process and we did not develop the polarization spectrum.}. Also, we used a standard parametrization of ${\cal R}_q^0$ given by

\begin{equation}
  |{\cal R}_q^0|^2=N^2q^{-3}\left(\frac{q/R_0}{\kappa_{\cal R}}\right)^{n_s-1},
\end{equation}

where $n_s$ is the spectral index. It is usual to take $\kappa_{\cal R}=0.05\;\text{Mpc}^{-1}$. 

All these definitions are consistent. Then, if we use $q=\beta l/r_{ls}$ we obtain 

\begin{eqnarray}
  |{\cal R}_{\beta l/r_{ls}}^0|^2&=&N^2\left(\frac{\beta l}{r_{ls}}\right)^{-3}\left(\frac{\beta l}{\kappa_{\cal R}r_{ls}}\right)^{n_s-1} =N^2\left(\frac{\beta l}{r_{ls}}\right)^{-3}\left(\frac{\beta l a_{DG}(t_{ls})}{\kappa_{\cal R}r_{ls}a_{DG}(t_{ls})}\right)^{n_s-1}\\&=&N^2\left(\frac{\beta l}{r_{ls}}\right)^{-3}\left(\frac{\beta l a_{DG}(t_{ls})}{\kappa_{\cal R}d_A(t_{ls})}\right)^{n_s-1}\equiv N^2\left(\frac{\beta l}{r_{ls}}\right)^{-3}\left(\frac{\beta l}{l_R}\right)^{n_s-1}.
  \label{R0_coeff}
\end{eqnarray}

Using similar calculations for the other distances, the final form of the Form Factors are given by

\begin{eqnarray}
{\cal F}(q)&=&\frac{{\cal R}_q^o}{5}\left[3{\cal T}(\beta l/l_T)R_{ls}-(1+R_{ls})^{-1/4}e^{-\beta^2l^2/l_D^2}{\cal S}(\beta l/l_T)\cos\left(\beta l/l_H+\Delta(\beta l/l_T)\right)\right]\,,
\label{Fq_definitivo}
\\
 {\cal G}(q)&=&\frac{\sqrt{3}{\cal R}_q^o}{5(1+R_{ls})^{3/4}}e^{-\beta^2l^2/l_D^2}{\cal S}(\beta l/l_T)\sin\left(\beta l/l_H+\Delta(\beta l/l_T)\right)\,,
 \label{Gq_definitivo}
\end{eqnarray}

where 

\begin{equation}
\label{l_i_coefficients}
l_R=\frac{\kappa_{{\cal R}}d_A^{DG}(t_{ls})}{a_{DG}(t_{ls})}\;,\;\;\; l_H=\frac{d_A^{DG}(t_{ls})}{d_H^{DG}(t_{ls})}\;,\;\;\; l_T=\frac{d_A^{DG}(t_{ls})}{d_T^{DG}(t_{ls})}\;,\;\;\;  l_D=\frac{d_A^{DG}(t_{ls})}{d_D^{DG}(t_{ls})}\;.
\end{equation}

To summarize, for reasonably large values of $l$, the CMB multipoles are given by 

\begin{eqnarray}\label{final multipole formula}
  \frac{l(l+1)C_{TT,l}^S}{2\pi}&=&\frac{4\pi T_0^2l^3\exp(-2\tau_{reion})}{r_{ls}^3}\int_1^{\infty}\frac{\beta d\beta}{\sqrt{\beta^2-1}}\nonumber\\
  &&\times\left[\left({\cal F}\left(\frac{l\beta}{r_{ls}}\right)+\tilde{{\cal F}}\left(\frac{l\beta}{r_{ls}}\right)\right)^2+\frac{\beta^2-1}{\beta^2}\left({\cal G}\left(\frac{l\beta}{r_{ls}}\right)+\tilde{{\cal G}}\left(\frac{l\beta}{r_{ls}}\right)\right)^2\right]\,.
\end{eqnarray}

We emphasize that the structure of the Equation \eqref{final multipole formula} considers that the Delta sector contributes additively inside the integral. If we set all Delta sector equal to zero, we recover the result for the scalar temperature-temperature multipole coefficients in GR given by Weinberg \cite{weinberg2008cosmology}.  Thus, the Equation \eqref{final multipole formula} is the main expression to implement the numerical analysis.\medskip

The DG contribution appears in many different forms in the Equation \eqref{final multipole formula}. The most notorious contribution is given by the functions $\tilde{{\cal F}}$ and $\tilde{{\cal G}}$. These functions are given by the functions $f,r,d,g$ and
$\tilde{f},\tilde{r},\tilde{d},\tilde{g}$ through the Equations \eqref{Gr eq 1} - \eqref{Gr eq 2}, and \eqref{Dg eq 1} - \eqref{Dg eq 2}. They are related to the evolution of the perturbation, and all these functions are coupled with the GR solutions.\medskip

The standard way to solve this problem is to obtain an analytical solution for the approximated Equations \eqref{GR pert solutions 1} - \eqref{GR pert solutions 3}, and solve them for every $\kappa$ (for example, from 0 to 100). Finally, match both results numerically, and solve $T, S$ and $\Delta$ as a function of $\kappa$. These equations evolve the perturbations given by the $f,r,d,g$ and
$\tilde{f},\tilde{r},\tilde{d},\tilde{g}$ functions, and then they must be evaluated inside the matter regime. They start to evolve inside the matter-era, but very close to the radiation era. This parametrization is given by $y = a/a_{EQ}$. The solutions were obtained starting from $y < 10^{-4}$ and stopping at $y \approx 10^{2}$. If the solutions are evaluated after the equality time, they could change, but, they are stable after $y\approx 10^2$.\medskip

\begin{figure}
\centering
 \includegraphics[width=.4\linewidth]{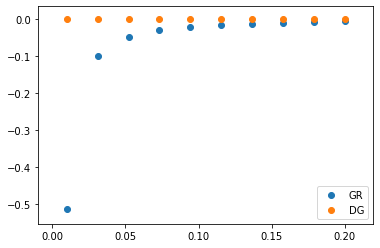}
 \caption{Comparison between the Form Factor ${\cal F}$ for GR (blue) and $\tilde{{\cal F}}$ for DG (orange).}
 \label{fig:F_formfactors}
\end{figure}

\begin{figure}
  \centering
  \includegraphics[width=.4\linewidth]{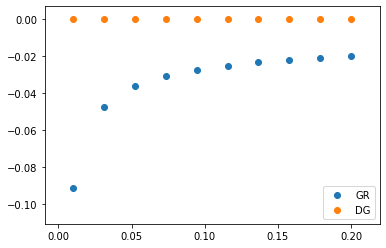}
\caption{Comparison between the Form Factor ${\cal G}$ for GR (blue) and $\tilde{{\cal G}}$ for DG (orange).}
  \label{fig:G_formfactors}
\end{figure}

The TT CMB spectrum needs these solutions because they build the Form Factors, and they are evaluated in an arbitrary $\kappa$ that is related to $\beta$ and $l$ through the Equation \eqref{final multipole formula}.  First, we found the results for the numerical solutions of $f,r,d,g$ and
$\tilde{f},\tilde{r},\tilde{d},\tilde{g}$, and then solve the expressions $T,S$ and $\Delta$. Then we calculate the Delta perturbations, and finally we obtain the Delta Form Factors. The Figures \ref{fig:F_formfactors} and \ref{fig:G_formfactors} shows the Form Factors for the background ${\cal F}$ and ${\cal G}$ and for the Delta contribution: $\tilde{{\cal F}}$ and $\tilde{{\cal G}}$. The Delta contribution is negligible with respect to ${\cal F}$ and ${\cal G}$. Numerically, the Delta contribution is $\approx 10^{39}$ times smaller than the Common Form Factors, thus we neglect the $\tilde{{\cal F}}$ and $\tilde{{\cal G}}$ terms.\medskip

However, the DG contribution appears in other ways. The next stage is going to be divided in three parts. The first is about the $l_i$ factors, the physics behind them, and the dependencies with physical processes. This is the biggest constraint that DG has. The second part is about the algorithm to include all the physical effects and the equations to obtain the TT CMB spectrum. The third and final part is about the results.

\subsubsection{\texorpdfstring{$l_R$}{lR}}

This coefficient depends on the angular distance and the DG Scale Factor $a_{DG}$ evaluated at the Last Scattering time. This term is associated with the $\mathcal{F}$ and $\mathcal{G}$ functions and depends on $n_s$, the spectral index of the primordial spectrum. In the case where the contribution to the Delta Form Factors is $\sim 0$, then the coefficient given by the Equation \eqref{R0_coeff} appears as a number powered to $n_s-1$. This factor appears in the Equation \eqref{final multipole formula} in front of the integral and regulates all the spectrum amplitude. We decided to assume an arbitrary $n_s$ to include the $l_R$ coefficient. This assumption is important because, at first glance, these parameters appear to be correlated: $N$, $n_s$ and $l_R$. This idea is incorrect because the $l_R$ value depends on the Last Scattering moment, defined by $z_{ls}$, and this redshift appears in many other places of the Equation \eqref{final multipole formula}. If $z_{ls}$ is not arbitrary, then the coefficient in the Equation \eqref{R0_coeff} is unique, and then $N^2$ have to compensate for the scale of the spectrum to fit the observable data. The $l_R$ parameter is a function of $z_{ls}$ and $ C $.\medskip

\subsubsection{\texorpdfstring{$l_H$}{lH}}

We followed the notation introduced in \cite{weinberg2008cosmology}, but the most known notation is $\theta = 1/l_H$.
If we want to preserve the CMB TT spectrum, we must use a value close to the standard $\theta$, but not strictly the same.  In this context, it is essential to remember that in the SNe-Ia analysis, we worked with $C=0$. This implies that there is no radiation and it is contradictory to the CMB procedure. Nonetheless, the SNe-Ia analysis is compatible with $C$ small values. Then, we can try to fit the TT CMB spectrum assuming a small $ C $ value, where $M \approx -19.3$ and the $H_0$ local value is preserved. We are going to work only in this scenario. Then, the CMB fit assumes a fixed $L_2$ value from SNe-Ia (we do not want to change this value) and a $C$ value close to $0$. After this process, we have to check that the $C$ value found by this method is compatible with the SNe-Ia data.\medskip

The most notorious constraint from the CMB spectrum is the acoustic peak position. This parameter determines the TT CMB spectrum (in the $l$ scale) and fits the hydrodynamic approach to the $l$-axis. Also, another important property of $\theta$ is that is obtained directly from the CMB spectrum. It's not a derived parameter \cite{Aghanim:2018eyx}:

\begin{equation}
     100\theta_{Planck}=1.0411\pm 0.0003.
\end{equation}

This value almost always appears in the literature as $\theta_{MC}$, where it was obtained by fitting the CMB data. However, in this work we calculate $l_H = 1/\theta$ as a function of $d_H^{DG}$ and $d_A^{DG}$. In our case, $\theta$ is not constraining the peak position by itself, we are constraining the $z_{ls}$, $ C $, and $h^2\Omega_{b,0}^{DG}$ values.

The physical meaning of this parameter is: the angle that subtends the size of fluctuation respect to the distance to this fluctuation. $d_H^{DG}$ is the horizon distance (size of the Universe at a specific redshift given by when the photons were decoupled). $d_A^{DG}$ is the angular distance between us and the TT CMB fluctuation. This relation must be corrected changing the speed of light $c$ by $c_s$ (the speed of sound) because it is the growing fluctuation speed  (\cite{Planck2015,Planck2018}). The correction has been introduced in Equations \eqref{HorizonDistanceDG} and \eqref{speed_equation_DG}.\medskip

The Fourier modes give an easy way to understand the dependence between $\theta$ and $l$. For simplicity, in a flat Universe, the modes of wavelength $\lambda \sim 2\pi a(t_{ls})/k$ on the Last Scattering surface seen today
under an angle $\theta = \lambda/d_A(t_{ls}) \sim 2\pi /l$ (the factor 2 appears because for a given multipole,
$\pi/l$ gives the angle between a maximum and a minimum. This is half of the wavelength of the perturbation on the surface). \cite[p.~228]{lesgourgues_mangano_miele_pastor_2013}  
This position of the peak is very well determined; then, this parameter is very well constrained. This condition imposes constraints over $C$ or $z_{ls}$ or $c_s$ (the speed of sound in a specific period: from $z=\infty$ to $z_{ls}$). In this analysis $L_2$ is fixed, and is independent of any other value that we are changing.\medskip

From the Equation \eqref{c_s} and knowing $R$, we can obtain the $d_H(z)^{DG}$ value in order to calculate $\theta$. As we have seen, $R$ is the baryons-photons relation. This factor considers particles that interact with the fluid, and then, the physical phenomena are described as sound waves. We can change this parameter if we suppose that more components interact in the fluid. But, we assume only the case where the photon-baryon relation determines the horizon distance.\medskip

The $R$ relation to calculate the speed of sound, is determined with $h^2\Omega_{b,0}^{DG}$ and $h^2\Omega_{\gamma,0}^{DG}$ values. This is very important because these parameters are physical and not apparent magnitudes. First of all, they depend on $Y_{DG}$ and not directly on $Y$. Second, they are physical magnitudes, they represent the real density of energy per volume, and then the interactions determine the physical speed of sound.\medskip

The CMB radiation gives physical density of photons: the blackbody spectrum has associated the $T_0$ temperature, where the real density is described as $\rho_{r,0} \propto T^4_0$ (Stefan-Boltzmann law). We know that the real physical densities in DG evolve with $Y_{DG}$, then it is easy to evolve any physical parameter as a function of $Y_{DG}$\footnote{Note: the parameters $h^2\Omega_{i,0}$ does not depend on $H$ or any other cosmological parameters. They are pure physical densities because of the critical density definition.}. The $l_H$ parameter is a function of $z_{ls},C$ and $h^2\Omega_{b,0}^{DG}$.

\subsubsection{\texorpdfstring{$l_T$}{lT}}

The $l_T$ parameter appears also inside of $\cos$ and $\sin$ functions in Equations \eqref{Fq_definitivo} and \eqref{Gq_definitivo}. Nevertheless, they move the $\cos$ and $\sin$ on the horizontal axis through the $\Delta$ Transfer function. They also appear outside the sinusoidal solutions, regulating the amplitude of these oscillations. The role of these parameters is to convert the arguments of the Transfer functions into the correct units. The origin of this normalization comes from the Equations \eqref{kappa} and \eqref{kappa_dT_transformation}. Those definitions are important because it implies that $d_T \propto a_{DG}(t_{ls})$, where $z_{ls}$ determines the DG Scale Factor at the moment of the Last Scattering. This normalization of the wave-number appears until this step of the numerical evaluation.

To evaluate this function, first we solve $Y$ as function of $z_{ls}$, and then evaluates $a_{DG}(t_{ls})$. Finally, it returns $d_A^{DG}/d_T^{DG}$ for that particular combination of $z_{ls}$ and $C$. Remember that $l_T$ parameter modulates the position and the amplitude of the $\sin$ and $\cos$ functions. Thus it is not trivial to know if this parameter is degenerated with another. Also, this is the only parameter that appears as an argument for the Transfer functions. Then, the result depends on the numerical solution of the Transfer functions. The $\mathcal{T},\mathcal{S}$ and $\Delta$ functions, can be solved numerically from the differential equations given by Equations \eqref{Gr eq 1} - \eqref{Gr eq 2} and the $\mathcal{T},\mathcal{S}$ and $\Delta$ definitions. The $l_T$ parameter is a function of $z_{ls}$ and $ C $.

\subsubsection{\texorpdfstring{$l_D$}{lD}}

Finally, the fourth parameter includes many steps that are related with physical processes. This parameter appears as a result of the physical damping of the oscillations, which is related to both processes: Silk and Landau dampings. These effects only appear next to every $\cos$ and $\sin$ function in the Equation \eqref{final multipole formula} as an exponential.
The TT CMB spectrum is very sensitive to this value because it changes the whole spectrum's amplitude.\medskip

First, the Silk damping is described by a special-relativistic non-perfect fluid. This approximation implies damping. The cosmology part appears when the damping effect acts on a range of time, and the effect must be integrated and corrected by the expanding Universe. The expression that describes the Silk damping is the Equation \eqref{Silk_damp}, where the cosmological correction appears with $Y_{DG}$.\medskip

Second, the calculation of Landau damping is challenging. Despite the Equation \eqref{Landau_damp} is very short, its intrinsic relation with the dispersion of the temperature creates many calculations. $\sigma_T$  is the standard deviation of the temperature at the Last Scattering moment when the transparency is a normal distribution function centered around the $z_{ls}$. This is a good approximation, but it requires many calculations provided by interactions related to the free electrons and photons.
In terms of the dispersion,

\begin{equation}
\sigma_t =\frac{\sigma_T}{T H_{DG}} ,
\end{equation}

because,

\begin{equation*}
\sigma_t dt = \sigma_T dT \to \frac{dt}{dT} = \frac{dt}{dY} \frac{dY}{dY_{DG}}\frac{dY_{DG}}{dT} \to \frac{dt}{dT} =  \frac{1}{H_{DG}T}
\end{equation*}

With this transformation, we can express the time-dispersion in terms of temperature. To obtain the temperature dispersion, first, we have to find the visibility function in DG, and before that, we have to define the Opacity function. This function is described in by \cite[~125p.]{weinberg2008cosmology} as

\begin{equation}
\mathcal{O}(T) = 1 - exp\left(-\int_{t(T)}^{t_0} c \sigma_\text{Thomson} n_e(t) dt\right).
\label{opacity_function}
\end{equation}

Another essential physical definition is the visibility function given by $O'(T)$, which describes the probability that the last scattering of a photon was at a temperature between $T$ and $T-dT$. It behaves like a probability distribution, then we try to find a normal distribution and obtain an estimation of $\sigma_T$ using the visibility function calculated $O'(T)$.

\begin{equation}
\mathcal{O'}_{fit}(T) \approx \frac{1}{\sigma_T \sqrt{2\pi }} e^{-{\frac {(T-T_L )^{2}}{2\sigma_T ^{2}}}}.
\label{normal_distribution}
\end{equation}

To obtain the $\sigma_T$ value, we evaluated the maximum of the distribution, where the $O'(T_{max})  \approx \frac{1}{\sigma_T \sqrt{2\pi }}$. \medskip

To calculate the opacity function, we have to know the physical electron density at that epoch. This is strictly related to the H, $e^-$, and $p$  abundances at that moment. These values can be easily correlated using an equation that describes the formation of the H. There are many methods to do this calculation. The most naive approximation is assuming an equilibrium through the Saha Equation. The equilibrium involves only atomic parameters, and it does not depend on cosmological parameters. Then, any assumption and equation in this calculation is preserved in DG. We emphasize that the evolution is given in terms of $T$.
Furthermore, the relation between $T$ and $z$ in DG is the same as in GR. Then, this procedure is totally preserved. In order to clarify any doubt, we are going to show the general scheme.\medskip

The naive approximation \cite[p.~113]{weinberg2008cosmology} begins at a time early enough so that protons, electrons, hydrogen, and helium atoms were in thermal equilibrium at the radiation's temperature. Then, the number density of any non-relativistic non-degenerate particle of type $i$ is given by the Maxwell-Boltzmann distribution:

\begin{equation}
    n_i = \frac{g_i}{(2\pi \hbar)^3} e^{\frac{\mu_i}{k_B T}} \int d^3q e^{-\frac{\left(m_i + \frac{q^2}{2m_i}\right)}{k_B T}}
    \label{MBdistribution}
\end{equation}

where $m_i$ is the particle mass, $g_i$ is the number of its spin states, and $\mu_i$ is the chemical potential of particles of type $i$. $g_p=g_e = 2$ while the 1s ground state of the H has two hyperfine states with spins $0$ and $1$, so $g_{1s}=1+3=4$. The most dominant reaction is given by $p+e \rightleftarrows H_{1s}$. The equilibrium is described by 

\begin{equation}
    \mu_p + \mu_e \rightleftarrows \mu_{1s}.
\end{equation}

Then, the relation between the density numbers is described by

\begin{equation}
    \frac{n_{1s}}{n_p n_e} = \left(\frac{m_ek_B T}{2\pi \hbar^2}\right)^{-3/2} e^{\frac{B_1}{k_BT}},
\end{equation}

where $B_1 \equiv m_p +m_e - m_H = 13.6 $ eV is the binding energy of the 1s ground state of the hydrogen. Now, including that $n_e = n_p$ because the Universe has to be neutral, and also consider that 76\% of the baryons were neutral or ionized hydrogen: $n_p+n_{1s} = 0.76 n_B$ \cite[p.~114]{weinberg2008cosmology}, we can define the fractional hydrogen ionization as $X\equiv n_p/(n_p+n_{1s})$, where the Saha equation is satisfied as:

\begin{equation}
    X(1+SX) = 1.
\end{equation}

Finally, $S$ can be expressed as

\begin{equation}
    S = \frac{(n_p+n_{1s})n_{1s}}{n_p^2} = 0.76 n_B \left(\frac{m_e k_B T}{2\pi \hbar^2}\right)^{-3/2} e^{B_1/k_B T}.
\end{equation}

Note that $S$ can be expressed in terms of $T$ and $h^2\Omega_{b,0}^{DG}$ as

\begin{equation}
    S = 1.747 \times 10^{-22} e^{157894/T} T^{3/2} h^2\Omega_{b,0}^{DG}.
    \label{naive_app}
\end{equation}

This dependence is significant for DG. First of all, the evolution is in terms of $T$ and not cosmic time, and also, the fraction $S$ depends on the baryon density parameter $h^2\Omega_{b,0}^{DG}$, then it will appear as a free parameter in the TT CMB spectrum. In DG, as we have said, the effect of Delta fields does not affect the spectrum (they are minimal). Only the evolution in time, represented by distances, can be affected by DG.

To improve the calculation, it is possible to add more corrections, including the 2$p$ and 2$s$ levels of the H atom. The full discussion about the decay and the emission processes can be found in \cite[p.~116]{weinberg2008cosmology}. 

The differential equation that describes this process with all those corrections is given by

\begin{equation}
    \frac{dX}{dT} = \frac{\alpha n}{H^{DG}T} \left(1+ \frac{\beta}{\Gamma_{2s}+\frac{8 \pi H^{DG}}{\lambda^3_\alpha n (1-X)}}\right)^{-1}\left(X^2-
    \frac{1-X}{S}\right),
    \label{dXdT_2s2p}
\end{equation}
where $n = n(h^2\Omega_{b,0}^{DG},T)$, $H^{DG} = H^{DG}(C,L_2,Y(T))$, and $\alpha = \alpha(T)$, $\beta = \beta(T)$ are functions related to the transitions of the H \footnote{For more details see
 \cite{weinberg2008cosmology}.}. This equation depends on the Hubble parameter: $H^{DG}$. This is important because in the derivation of this equation, $H^{DG}$ appears in two different places: the first term $1/TH^{DG}$ is a coefficient that comes from changing $t$ to $T$ (to evolve the equations in temperature instead of time) and the second term (where $H^{DG}$ appears as $8\pi H^{DG}$) comes from the change of the frequency (or wavelength) produced by the cosmic expansion. Therefore, both of those corrections appear in DG as $H^{DG}$ and not like the standard $ H $  (then, this equation looks similar, but it is different because the dependence between the variables is totally different) \cite[p. 122]{weinberg2008cosmology}.\medskip
 
 In DG, this effect could be crucial because the evolution could change due to that the Hubble parameter is a function of the Effective Scale Factor $Y_{DG}$, and this is a function of $Y(t)$. Furthermore, the $T$ preserves the standard dependence with the Effective Scale Factor $Y_{DG}$, in other words, in standard cosmology, we have $T=T_0(1+z)$ and this relation is preserved in DG, but the dependence between $z$ in DG appears related to $a_{DG}(Y(t))$. Furthermore, the numerical solution with all these corrections changes the Saha approximation, and then also changes the GR solution. It is also essential to note that the differential equations are evolved in a high range of $T$, and DG tends to be very similar to the standard GR at the beginning. The Scale Factor tends to be the same because the Delta field contributions disappear when $Y\to 0$. Nevertheless, all these aspects must be taken into account to compute $X(T)$ in order to obtain an excellent numerical value to fix $z_{ls}$ and $n_e$ affecting the Visibility function: the peak position in redshift ($z_{ls}$) and the standard deviation ($\sigma_T$).
We remark that the $\alpha(T)$ and $\beta(T)$ are numerical functions of $T$ \cite{alpha_coefficient}  and there is no cosmological influence here, then it does not affect the DG calculations. The Visibility is a function of $C$ and $h^2\Omega_{b,0}^{DG}$.  This function is essential to find the $z_{ls}$ because the peak is associated to the $z_{ls}$. The $l_D$ parameter is a function of $z_{ls}, C$ and $h^2\Omega^{DG}_{b,0}$.

\subsubsection{Algorithm to obtain the CMB}

The MCMC algorithm consists of a modified Adaptative Metropolis MCMC algorithm.We used the TT spectrum from \cite{Planck2018}. \footnote{The data were obtained from \url{https://pla.esac.esa.int/\#cosmology}.} 

In our case, we want to find all the possible values that match, in the best way, the TT CMB spectrum. The algorithm works as follows: we propose an original distribution of values, called priors: $C,h^2\Omega_{b,0}^{DG},z_{ls},n_s$ and $N$. which are all normally distributed. Then we calculate the predicted TT CMB spectrum and comparing with the TT CMB spectrum from \cite{Planck2018}. The likelihood is defined as usual, based on the squared error.

We introduced a modification to give more flexibility in the $z_{ls}$ fitting. We did not want to force the system to choose a $z_{ls}$ exactly in the peak postion of the visibility function, therefore we create a proposal distribution centered in the peak of the visibility function, and then, then MCMC takes that prior and move it around the peak. With this method, we give more freedom to the $z_{ls}$ parameter and the final posterior probability associated to this parameter could be slightly different from the peak of $O'(T)$. All the others parameters were found as the standard Metropolis MCMC.\medskip

\section{Results}

First of all,  we clarify that all the chains always converged to the same values; all are independent of the prior distributions. Now, we present the results. This corresponds to a chain with 20.000 steps for every parameter.\medskip 

\begin{figure}[h!]
\centering
\includegraphics[width=16cm,keepaspectratio]{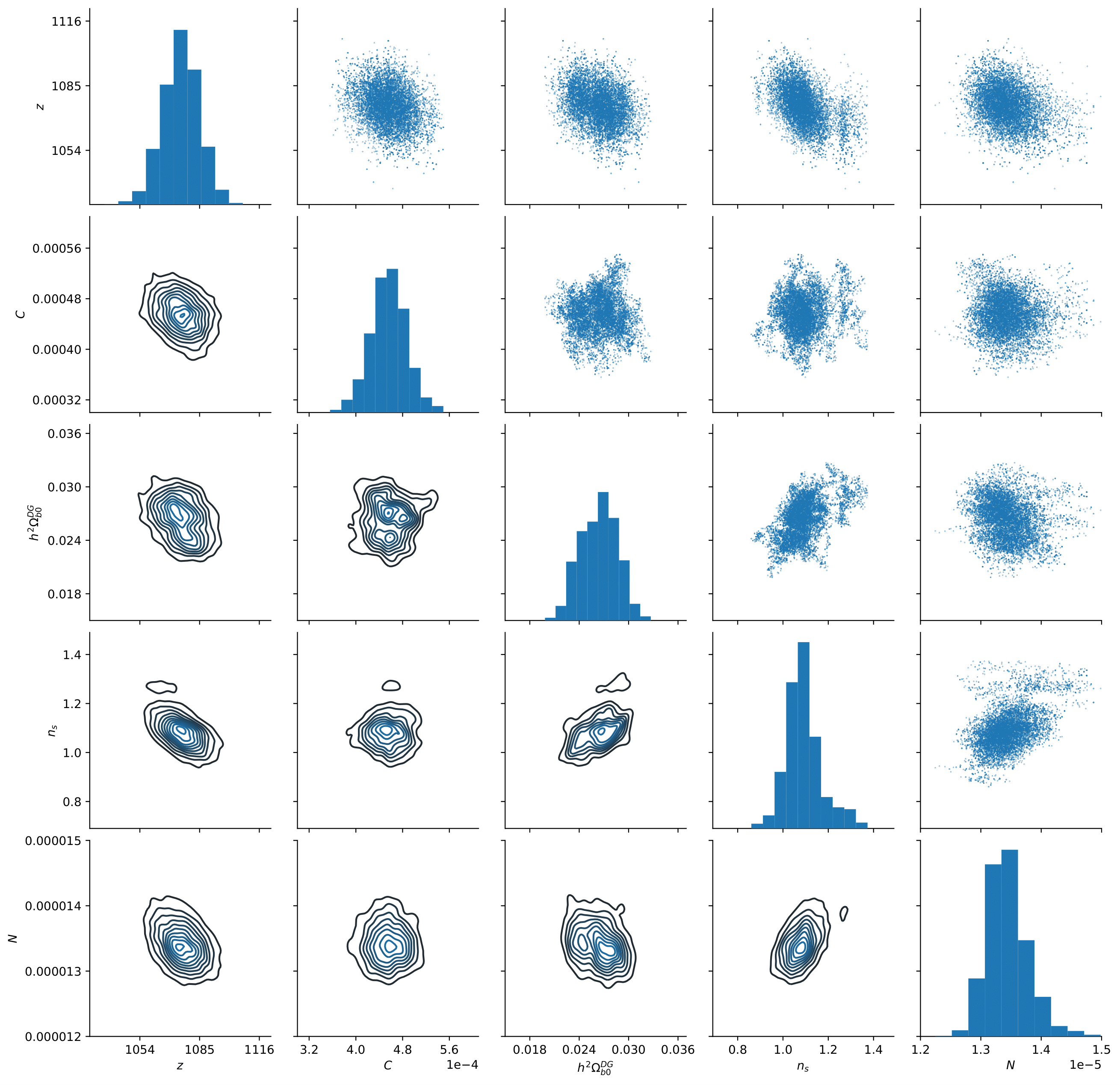} 
\caption{Contour plot for all posterior probabilities associated to the DG parameters.}
\label{fig:full_MCMC_CMB}
\end{figure}

The posterior distribution for every parameter are shown in the Figure \ref{fig:full_MCMC_CMB} in the diagonal. All the distributions show only one peak, but some of them are not normally distributed. We specify the case of ${h^2\Omega_{b,0}}^{DG}$ and $n_s$. These parameters show multimodal distributions but always with a clearly main peak. We fit in both cases a normal distribution but the error was defined such that the $\sigma_x$ includes the smallest multimodal distributions with its errors. Then, all the parameters have errors defined as $\pm 1 \sigma_x$, with exception of the baryon density parameter which is ${h^2\Omega_{b,0}^{DG}}_{-2\sigma}^{+2\sigma}$ and the spectral index given by  ${n_s}_{-2\sigma}^{+3\sigma}$.\medskip

\begin{table}[h]
\caption{MCMC fit results for the DG free parameters. These values are related to posterior distributions.} 
\centering
\begin{tabular}{|c|c|c|}
\hline
Parameter              & Mean                 & Standard deviation   \\ \hline
$z_{ls}$               & $1075.3$             & $9.4$                \\ \hline
$C$                    & $4.6 \times 10^{-4}$ & $0.3\times 10^{-4}$  \\ \hline
$h^2\Omega_{b,0}^{DG}$ & $0.026$              & $0.002$              \\ \hline
$n_s$                  & $1.09$               & $0.08$               \\ \hline
$N$                    & $1.34\times 10^{-5}$ & $0.04\times 10^{-5}$ \\ \hline
\end{tabular}
\label{Tab:MCMCresultsCMB}
\end{table}

\begin{figure}[h!]
\centering
\includegraphics[width=16cm,keepaspectratio]{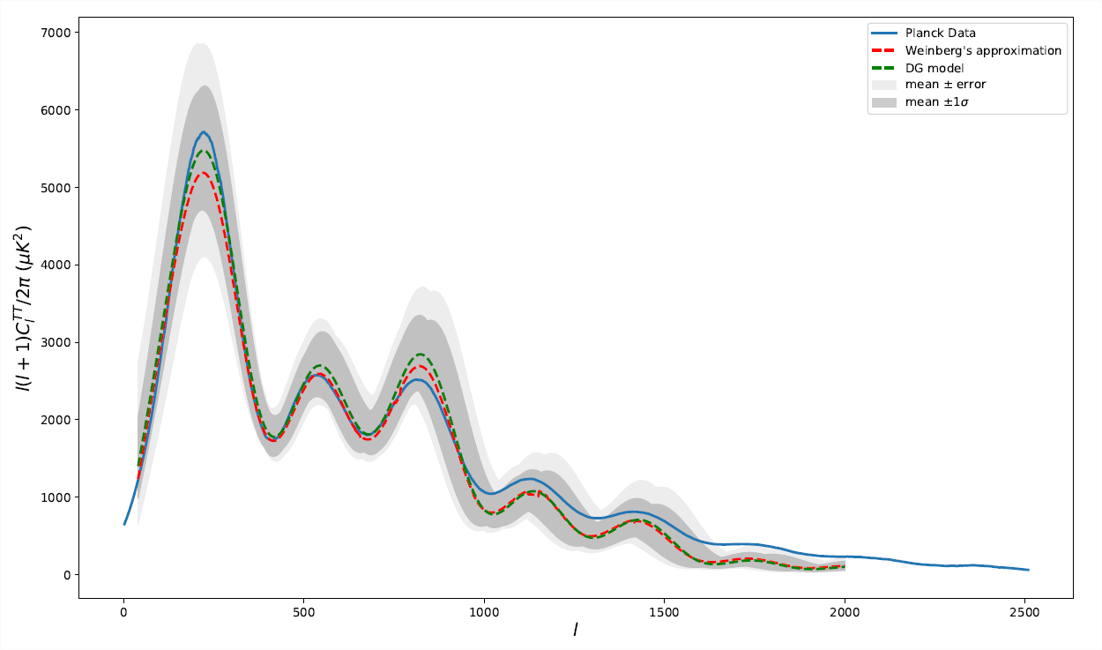} 
\caption{TT CMB spectrum was predicted by DG vs. the observed TT CMB spectrum. The blue line corresponds to the Planck observations, the green line is the DG prediction, and the greyscale is the error associated with the MCMC posterior probabilities. The red line is the solution obtained with the Weinberg's approximation, assuming the standard $\Lambda$CDM cosmology  (\cite{weinberg2008cosmology}).}
\label{fig:final_fit}
\end{figure}

The Figure \ref{fig:full_MCMC_CMB} shows all the combinations for the 5 free parameters. All the parameters are constrained to a normal-like distribution, and they are independent of each other.
Then, the shape of the TT CMB spectrum constraint all the parameters to ``accurate'' values. The fitted curve is shown in the Figure \ref{fig:final_fit}.\medskip

These results are good according to the approximation given by \cite{weinberg2008cosmology}. This analytic and hydrodynamic approach shows a good fit for the most prominent three peaks, including the acoustic peak, but it is inaccurate at larger multipoles. The Figure \ref{fig:final_fit} shows that DG prediction is very similar to the observable data, but the prediction is inaccurate from the third peak on. However, the precision of the approximation includes that error scale. In \cite{weinberg2008cosmology} the TT CMB spectrum has a similar error, and the differences also appear at larger multipoles. The DG TT CMB spectrum (green line) calculated with this approach is very similar to the spectrum calculated with $\Lambda$CDM model (red line).\medskip

Two important aspects must be checked: the $ C $ value and the Visibility function peak compatibility with the $z_{ls}$ needed to fit the TT CMB spectrum.\medskip

Respect to the $C$ value, the TT CMB spectrum fix this value around $C = 4.6\times 10^{-4}$. This result is completely in concordance with SNe-Ia results. The $C$ parameter is so small that the SNe-Ia analysis cannot detect a difference between $0$ and $\approx 10^{-4}$. Then, the $M$ and $H_0$ observables obtained from \cite{Riess2016,Riess2018,Riess2019} are in concordance with our results, assuming a standard error in the approximation of the hydrodynamic approach similar to GR.\medskip

\begin{figure}[h!]
\centering
\includegraphics[width=16cm,keepaspectratio]{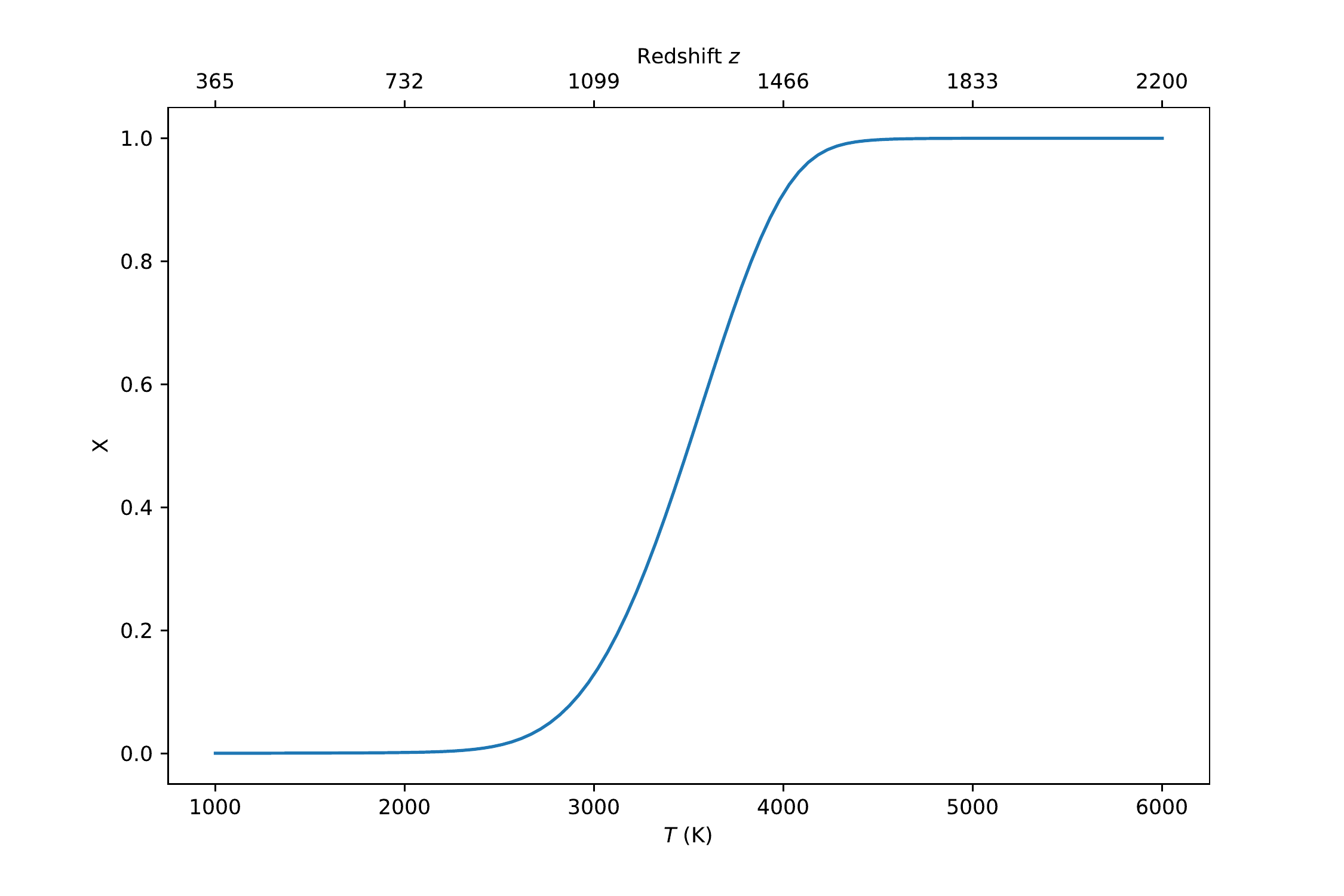} 
\caption{$X(T)$ fraction as function of temperature $T$ and redshift $z$ assuming $C$ and $h^2\Omega_{b,0}^{DG}$ MCMC results.}
\label{fig:X_de_T}
\end{figure}

In the Last Scattering redshift case, we have to check if $z_{ls}$ is close to the Visibility function peak. The Figure \ref{fig:X_de_T} shows how the fraction of free electrons $X$ depends on $T$ and $z$. At lower temperatures $X \to 0$, meanwhile at higher temperatures $X \to 1$. The $ X $ function depends on $C$, $h^2\Omega_{b,0}^{DG}$ and $T$, where the MCMC results have fixed the two first parameters. This case is shown in the Figure \ref{fig:X_de_T}.\medskip

\begin{figure}[h!]
\centering
\includegraphics[width=16cm,keepaspectratio]{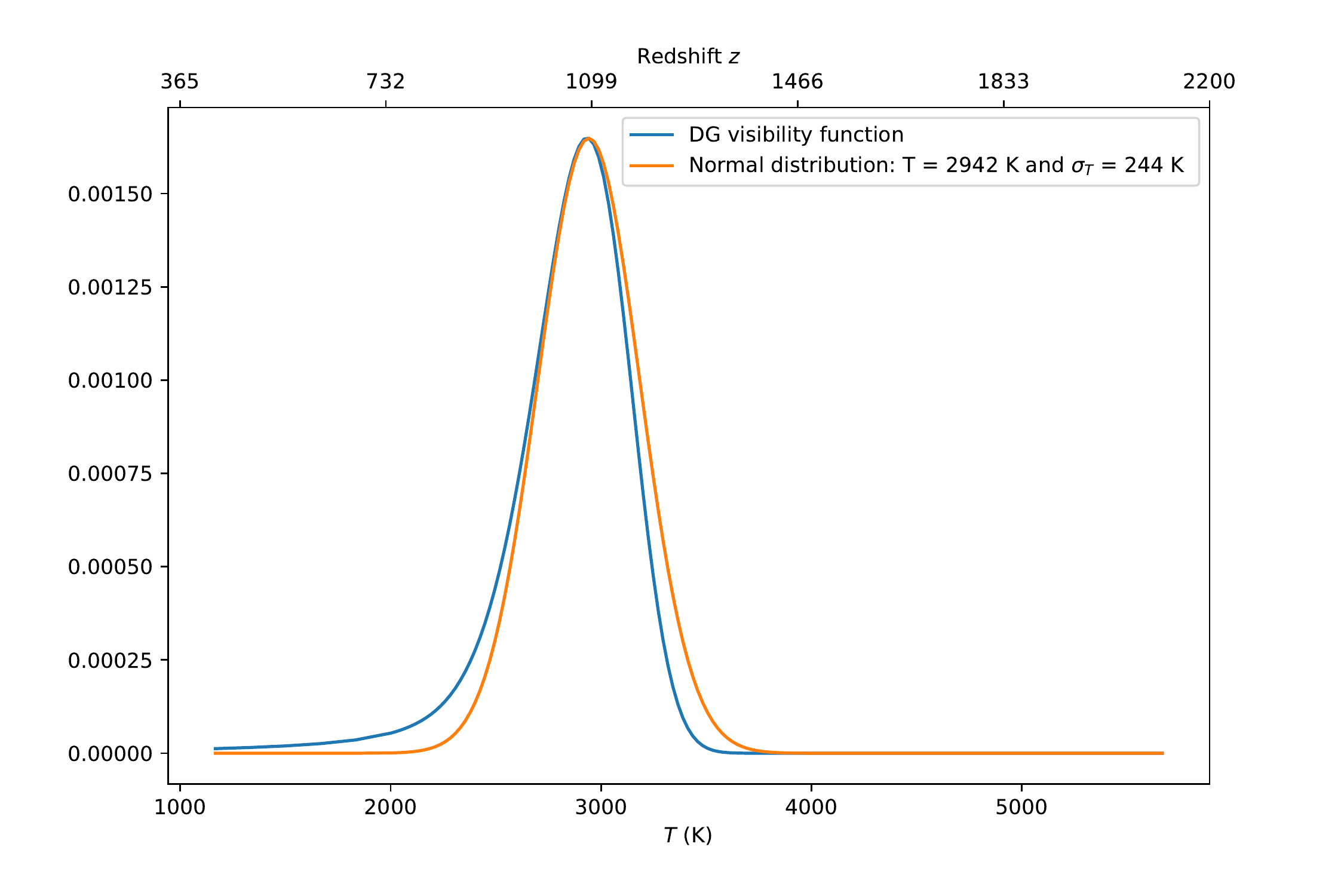} 
\caption{In blue color, the Visibility function is associated with the $X(T)$ obtained from the MCMC results. The orange line is a normal distribution centered in the peak of the DG solution.}
\label{fig:vis_fun}
\end{figure}

Then, the visibility function has a maximum close to $T_{max} \approx 2942$ K ($z_{max}\approx 1078$) with a temperature dispersion $\sigma_T \approx 244$ K. This function is shown in the Figure \ref{fig:vis_fun}. Furthermore, we add a normal distribution centered at the same peak to show the similarity between the Visibility function and a normal distribution. 

The $\sigma_T$ was estimated from the height of the peak (not by fitting a distribution, FWHM, or any other method). 

The GR case \cite{weinberg2008cosmology} finds $T_{max} \approx 2941$ K with a $\sigma_T \approx 248$ K. While, the DG peak around $z \approx 1078$ is near the MCMC results $z_{ls} \approx 1075$. Despite $z_{ls}$ was obtained varying the redshift around the peak estimation, the $z_{ls}$ is not exactly the peak associated with the Visibility function, but it is near.\medskip

Finally, the density of matter and radiation is related to the $C$ and $L_2$ values through the definition of the physical densities. In GR, the equality moment is vital because the hydrodynamic approach uses equality to match the equations when the Universe was dominated by radiation and dominated by matter. In the case of GR, naturally appears that

\begin{eqnarray}
\frac{\rho_{GR,m}}{\rho_{GR,r}}=\frac{Y}{C}\;,
\end{eqnarray}

where $C=\Omega_{r,0}/\Omega_{m,0}$ by definition. Then the moment of equality in GR corresponds to $Y_{EQ}=C$. But, for DG densities, the physical densities depend on $Y_{DG}$, thus

\begin{eqnarray}
\frac{\rho_{DG,m}}{\rho_{DG,r}}=\frac{Y_{DG}}{C_{DG}}\;,
\end{eqnarray}

where $C_{DG}=\Omega^{DG}_{r,0}/\Omega^{DG}_{m,0}$. In DG, we imposed that the equality moment must occur in both sectors at the same time. In other words,

\begin{equation}
    Y_{DG}(Y_{EQ})=C_{DG} \to C_{DG}=C\frac{\sqrt{\frac{1+F(C)}{1+3F(C)}}}{\sqrt{\frac{1+F(1)}{1+3F(1)}}}\;,
\end{equation}

From the MCMC results, we know that $C\ll 1$ and $L_2 \approx 0.45$, then

\begin{equation}
C_{DG}\approx C\sqrt{\frac{1-L_2}{1-L_2/3}}.
\end{equation} 

This result is useful because if we know the physical density of radiation, we can find the physical density of matter. Then,

\begin{equation}
C_{DG}\approx C\sqrt{\frac{1-L_2}{1-L_2/3}} \approx 0.80 C \approx 3.7 \times 10^{-4}.
\end{equation} 

Note: To be clear, in the next calculations we emphasize the observable (physical) densities with a DG sub or superscript.\medskip

To calculate the physical densities, we can use the photon density given by the black body spectrum integrated (based on the TT CMB spectrum):

\begin{equation}
   \rho_{\gamma,0}^{DG} c^2=a_B T_0^4,
\end{equation},

where

\begin{equation}
   a_B=\frac{8\pi^5 k_B^4}{15 h^3 c^3}=7.56577 \times 10^{-16} \text{  J} \text{  m}^{-3} \text{  K}^{-4},
\end{equation}

is the radiation energy constant. With $T_0=2.7255 K$, we get the today density associated to the photons
$\rho_{\gamma,0 }^{DG}=a_B T^4_0 /c^2=4.64511\times 10^{-31} \text{kg m}^{3}$. This is a physical quantity.\medskip

The neutrinos density (physical quantity) is related to the photon density as following \cite{BIB_PLANCK_Ade:2013zuv} 

\begin{equation}
\rho_{\nu,0}^{DG} =N_{\text{eff}} \frac{7}{8}\left(\frac{4}{11}\right)^{4/3}\rho _{\gamma,0}^{DG},
\label{Neffeq}
\end{equation}

where $N_\text{eff}^\text{Planck} = 3.04678$ (\cite{Planck2018}). The relation given by the Equation \eqref{Neffeq} is based on statistical mechanics: photons and neutrinos are in thermal equilibrium, but neutrinos are fermions and photons are bosons. Thus,

\begin{equation}
\rho_{\nu,0}^{DG}=3.21334\times 10^{-31} \text{  kg} \text{  m}^{-3},
\end{equation}

and the total radiation density (physical quantity) is given by

\begin{equation}
\rho_\text{r,0}^{DG}=\rho_{\gamma,0}^{DG} + \rho_{\nu,0}^{DG}= 7.85846 \times 10^{-31} \text{  kg} \text{  m}^{-3}.
\end{equation}

Until here, we have assumed that neutrinos are relativistic particles and contribute to the radiation density. We can also write these values divided by the critical density given by:

\begin{equation}
\rho_{c,0}=\frac{3H^2_0}{8\pi G}=1.87847 h^2\times 10^{-26} \text{  kg} \text{  m}^{-3},
\end{equation}

where the GR Hubble Constant have been expressed in terms of the dimensionless parameter h, where $H_0=100 h$ $\text{km s}^{-1} \text{Mpc}^{-1}$. \footnote{The $\rho_{c,0}$ is not a physical density. The physical critical density is $\rho_{c,0}^{DG} = \frac{3H_{DG,0}^2}{8\pi G} = 1.87847 h_{DG}^2\times 10^{-26} \text{  kg} \text{  m}^{-3} $, where $H_{DG,0} = 100 h_{DG}$. The numerical factor is exactly equal in both cases, then the results shown in Equations \eqref{fisden1}-\eqref{fisden4} do not change.} Therefore, the density parameters are (these are physical!, we emphasize that the $h$ constant is simplified, these parameters are independent of $h$.)

\begin{equation}
\begin{split}
h^2\Omega_{\gamma,0}^{DG} &= \dfrac{\rho_{\gamma,0}^{DG}}{\rho_{c,0}}h^2 = 2.47\times 10^{-5},\\
h^2\Omega_{\nu,0}^{DG} &= \dfrac{\rho_{\nu,0}^{DG}}{\rho_{c,0}}h^2 = 1.71\times 10^{-5},\\
h^2\Omega_{r,0}^{DG} &= h^2\Omega_{\gamma,0}^{DG} + h^2\Omega_{\nu,0}^{DG} = 4.18\times 10^{-5},
\end{split}  
\end{equation}

and (cdm is ``cold dark matter'')

\begin{equation}
    h^2\Omega_{m,0}^{DG} \equiv h^2\Omega_{b,0}^{DG} + h^2\Omega_{cdm,0}^{DG} +  (3 - N_{\text{eff}}) h^2\Omega_{r,0}^{DG} \approx h^2\Omega_{b,0}^{DG} + h^2\Omega_{cdm,0}^{DG},
\end{equation}

Finally, we assume that $N_{\text{eff}} = 3$ (we emphasize, again, that $h^2\Omega_{x,0}^{DG}$ quantities are not related with $H_0$. They are related only with the physical density and $3\times100^2/8\pi G$) the quantities are:

\begin{eqnarray}
h^2\Omega_{r,0}^{DG} = 4.18\times 10^{-5} \label{fisden1},\\
h^2\Omega_{b,0}^{DG} = 0.026,\\
h^2\Omega_{m,0}^{DG} = 0.113,\\
h^2\Omega_{cdm,0}^{DG} \equiv h^2\Omega_{m,0}^{DG} - h^2\Omega_{b,0}^{DG}  = 0.087.
\label{fisden4}
\end{eqnarray}

\section{Conclusions}

We have studied the cosmological implications for a modified gravity theory named Delta Gravity. The results from SNe-Ia analysis indicate that DG explains the accelerating expansion of the Universe without  $\Lambda$ or anything like ``Dark Energy''. The Delta Gravity equations naturally produce the acceleration. 
In this work we performed a fit to the SNe-Ia data considering three free parameters $M$, $C$ and $L_2$, finding that $C$ is not relevant if it is small enough. Also we found that $L_2 \approx 0.457$ and $h \approx 0.496$, where $h$ is not the Hubble constant. Regarding $L_2$, this parameter establishes the acceleration of the Universe and it is independent of $M$, where $M$ is degenerated with $h$. In this case, the Universe is accelerating as a result of $L_2>0$ and implying that a new kind of densities called Delta matter and radiation must exist. These can be associated with the new Delta fields. It is not clear if this Delta Composition is made of real particles, or not. However, we propose two different interpretations. The first is that the Universe only contains matter (baryonic and cold dark matter) and radiation where the Delta sector is only a geometric effect. The other scenario is that the Universe also contains Delta matter and Delta radiation because they are particles. In both scenarios, the Universe shows the same behavior, and it is accelerating, but the difference is that in the first case the Delta Sector could be invisible because the geometry provides the fundamental physics behind Delta Sector and not the particles. This is part of the interpretation, and for now, we cannot conclude more about this aspect.\medskip

Regarding the TT CMB Spectrum,  we used 5 free parameters to fit it: $C,h^2\Omega_{b,0}^{DG},z_{ls},n_s$ and $N$.

The first peak is very well determined in position and shape, but not the other two peaks. In the GR case, they tend to be modulated by the dark matter and baryon density (\cite{camb2}  \footnote{Any dependence can be easily verified with \url{https://camb.readthedocs.io/en/latest/CAMBdemo.html}). Specifically, the dependence of the peak's heights and its relative positions respect to the $h^2\Omega_x$.}. Nevertheless, in the hydrodynamic approach \cite{weinberg2008cosmology}, the dark matter evolution is assumed as dominant considering that all the gravitational potential is driven by dark matter. This approximation is useful because the equations are easy to solve, however it is not accurate according to \cite[p. 358]{weinberg2008cosmology}: this approach introduced 10\% errors or less in the GR case. In DG we used the same approximation and obtain a very similar result. Despite this approximation, the TT CMB spectrum is very well described, but the large multipoles show deviations from the observable data. It is vital to consider that the fitted values were obtained from an approximation called hydrodynamic approach, and then, the numerical values contain intrinsic errors associated with the approximations, then they are not accurate. Nonetheless, these values are very similar to the GR case.\medskip

The $z_{ls}$ obtained from the MCMC is compatible with the transition range showed in Figure \ref{fig:X_de_T}, and the peak of the Visibility function showed in Figure \ref{fig:vis_fun}. 
The amount of baryonic matter given by $h^2\Omega_{b,0}^{DG} = 0.026$ is close to the GR case: $0.022$. It is important to contrast this value with other measurements, especially because DG has a very different description of the Universe, where other equations that are different to GR, give the distances. Then, other observational constraints must be examined meticulously in order to conclude if DG fit those observations.

The parameters related to the primordial spectrum, $ A $ and $n_s$, are close to the standard values: the spectral index is close to $1$, and the amplitude is $\sim 10^{-5}$.

An assumption that is essential for all the CMB analysis is that the plasma fluid, which is described with the speed of sound $c_s$ within the horizon radius, is only affected by baryons and radiation. This aspect could indicate that Delta Components do not interact with Common radiation and matter, but it would be interesting to analyze all the changes that introduce a Delta sector that interacts with Common matter and radiation. This aspect may change many approximations and, then, could affect enormously the TT CMB spectrum. This could be part of future research.\medskip

The observable rate of expansion of the Universe in DG is given by $H_0^{DG}$. This parameter is determined by $L_2$ and $h$. In the context of the TT CMB analysis, if $C$ is very small, then the SNe-Ia observations can be compatible with the TT CMB spectrum. The results show that $C\sim 10^{-4}$. In this regime, the SNe-Ia is not affected, and the compatibility between both observations is possible. It is important to emphasize that there are two values that are different. One is $h$, which is provided from the GR background, and second, the $H^{DG}_0$, that is the observable Hubble Constant in this model.\medskip

A relevant cosmological value that can be constrained from the observations, is the age of the Universe. The higher the Hubble Constant, the lower the age of the Universe. This relation is vital since if the local fit of supernovae radically changes $H_0$, then the age of the Universe changes. Therefore, there could be conflicts with some estimates of the age of the Universe that are independent of cosmology. We remark the fact that according to local measurements of supernovae, the age of the Universe for DG and GR are: 13.1 Gyrs for DG and  13.0 Gyrs for GR. Instead, Planck's data imply a larger age of the Universe: 13.8 Gyrs. A crucial and precise estimation based on the measurement of globular clusters age in the Milky Way \cite{refId0} \footnote{\url{https://www.eso.org/public/chile/news/eso0425/}}, which is independent of cosmology, indicates that the Universe has to be older than 13.6 $\pm$ 0.8 Gyrs. DG and GR, assuming the results of SNe's local measurements, are on the verge of this observational constraint. According to this, one wonders if SNe can be in conflict with the age of the Universe. It is a very recent discussion, and we are only commenting on the problems when astrophysicists try to make SNe and CMB compatible. We emphasize that the problem goes beyond DG because a high Hubble Constant causes it, and it also involves other types of measurements that yield high values of the Hubble Constant. This discrepancy could be caused by the calibrations and methods used by Riess et al., but this tension between both observations has been widely discussed and until now there is no agreement. Even, other researchers have tried to measure the $H_0$ value using methods independent of distance ladders and the CMB. They found that the Hubble Constant exceeds the Planck results, with the confidence of 95\% \cite{Pesce_2020}. However, other measurements based on the tip of the red giant branch (TRGB) have found that $H_0$ is close to $69.6 $ km/(Mpc s) \cite{Freedman_2019,Freedman_2020}. Other methods based on lensed quasars found that $H_0=73.3$ Mpc/(km s) agrees with local measurements but tension with Planck observations \cite{Wong_2020}.\medskip

All the TT CMB spectrum analysis was made in the DG context were the Delta contributions represented by $\tilde{F}$ and $\tilde{G}$ can be neglected. This is an essential part of the development of the perturbation theory, and it implied many simplifications when we want to calculate the spectrum and creates more constraints on the spectrum fitting.\medskip

To summarize, DG requires more development to compare with other constraints such as the He produced at the Big Bang nucleosynthesis, or the BAOs constraints, or even cosmological simulations. This last aspect could be relevant if the interpretation of the Delta Sector is given in terms of particles that create gravitational interactions. In fact, at the Newtonian limit, the Delta matter appears as a new source of the gravitational potential \cite{Alfaro:2019}.\medskip

Finally, it is remarkable that DG finds a well-behaved TT CMB spectrum, where it is possible to constraint new parameters, even related to inflation. However, this analysis does not use all the numerical precision, because the equations are only an approximation, and even more, we are calculating only the scalar contributions to the total TT CMB spectrum. Furthermore, many other sources that contribute to the ``spectrum'' have been avoided to simplify the analytical solution, such as the Sachs-Wolfe effect or lensing. This is only a first order approximation, and it shows that DG could fit the TT CMB spectrum, but it is essential to fit the spectrum with all the numerical precision without approximations because the conclusions drawn in that case could be different. Thus, these numerical results must be understood as values that are near the correct value, not as a final and undeniable result.\medskip

The incompatibility between the SNe-Ia and CMB occurs when $\Lambda$CDM model is constrained using BAOs and SNe-Ia. Even when the model uses curvature: if all the parameters describe the same Universe, the whole model must be compatible with only one geometry given by $\Omega_k$. For example, recently, it was published an article that shows a discrepancy between the Planck's data \cite{Planck2018}. These differences can be caused by the assumption that the Universe is flat. Despite this curvature assumption in the $\Lambda$CDM model, the cosmological parameters are incompatible because some of them are compatible with a flat Universe, but others indicate a closed Universe \cite{curvature}. Furthermore, regarding the SNe-Ia analysis, another article shows an anisotropy in the SNe-Ia distribution, and then, the acceleration measurement could be wrong \cite{anisotropy}.  All the DG analysis could change because the $L_2$ value will be different, and all the distances would change \cite{2019arXiv191204903K}. 
. In this context, it is relevant to emphasize that there are many approximations in our procedure, and DG must be contrasted with other observations to conclude with a good precision if this model is a solution for today's paradigm. BAOs could be an excellent option to verify the model, mainly because these observations are related to the angular distance and could constrain the DG model and verify if DG can survive to describe SNe-Ia and BAOs.  \medskip

Despite these interpretations, problems, and approximations, DG can fit both SNe-Ia and TT CMB spectrum data, without Dark Energy, but it is very necessary to include the complete numerical solutions without approximations to be able to conclude with certainty if DG can explain both phenomena.

\bibliography{sample63}{}
\bibliographystyle{aasjournal}

\end{document}